\documentclass{article}
\pdfoutput=1

\usepackage{epsf}
\usepackage{amsfonts}
\usepackage{amsmath}
\usepackage{amssymb}
\usepackage{booktabs}
\usepackage{epubtk}
\usepackage{graphicx}

\def\R{\mathbb R}

\newcounter{theorem}
\newenvironment{theorem}
  {\refstepcounter{theorem}
   \vspace{1em}
   \noindent{\bf Theorem~\thetheorem}
   \begin{em}}
  {\end{em}
   \newline}

\DeclareMathOperator{\diag}{diag}
\DeclareMathOperator{\tr}{tr}

\newcommand{\T}[1]{\textit{T}\super{#1}}

\begin{document}

\title{The Einstein--Vlasov System/Kinetic Theory}

\author{\epubtkAuthorData{H{\aa}kan Andr{\'e}asson}
                         {Mathematical Sciences\\
                         University of Gothenburg\\
                         Chalmers University of Technology\\
                         S-412\,96 G\"oteborg, Sweden}
                         {hand@math.chalmers.se}
                         {http://www.math.chalmers.se/~hand}
}

\date{}
\maketitle


\begin{abstract}
  The main purpose of this article is to provide a guide to theorems
  on global properties of solutions to the Einstein--Vlasov
  system. This system couples Einstein's equations to a kinetic matter
  model. Kinetic theory has been an important field of research during
  several decades in which the main focus has been on non-relativistic
  and special relativistic physics, i.e., to model the dynamics of
  neutral gases, plasmas, and Newtonian self-gravitating systems. In
  1990, Rendall and Rein initiated a mathematical study of the
  Einstein--Vlasov system. Since then many theorems on global
  properties of solutions to this system have been established.
  This paper gives introductions to kinetic theory in
  non-curved spacetimes and then the Einstein--Vlasov system is
  introduced. We believe that a good understanding of kinetic theory
  in non-curved spacetimes is fundamental to a good comprehension of
  kinetic theory in general relativity.
\end{abstract}

\epubtkKeywords{Einstein-Vlasov system, Global existence, Kinetic
  theory, Nordstr\"om-Vlasov system, Vlasov-Poisson system}

\newpage

\epubtkUpdate
    [Id=A,
     ApprovedBy=subjecteditor,
     AcceptDate={5 May 2011},
     PublishDate={27 May 2011},
     Type=major]{%
This is a revised and updated version of the article from 2005. A
number of new sections on the Einstein--Vlasov system have been added,
e.g., on the formation of black holes and trapped surfaces, on
self-similar solutions, on the structure of static solutions, on
Buchdahl type inequalities, on the stability of cosmological
solutions, and on axisymmetric solutions. Some of the previous
sections have been significantly extended. The number of references
has increased from 121 to 197.
}

\newpage
\tableofcontents

\newpage


\section{Introduction to Kinetic Theory}

In general relativity, kinetic theory has been used relatively
sparsely to model phenomenological matter in comparison to fluid
models, although interest has increased in recent years.
From a mathematical point of view there are
fundamental advantages to using a kinetic description. In non-curved
spacetimes kinetic theory has been studied intensively as a
mathematical subject during several decades, and it has also played an
important role from an engineering point of view. 

The main purpose
of this review paper is to discuss mathematical results for
the Einstein--Vlasov system. However, in the first part of
this introduction, we review kinetic theory in non-curved
spacetimes and focus on the special-relativistic case, although
some results in the non-relativistic case will also be
mentioned. The reason that we focus on the relativistic case is not 
only that it is more closely related to the main theme in this review, 
but also that the literature on relativistic kinetic theory is very sparse 
in comparison to the non-relativistic case, in particular concerning 
the relativistic and non-relativistic Boltzmann equation. 
We believe that a good understanding of kinetic theory in
non-curved spacetimes is fundamental to good comprehension of kinetic
theory in general relativity. Moreover, it is often the case that
mathematical methods used to treat the Einstein--Vlasov system are
carried over from methods developed in the special relativistic or
non-relativistic case.

The purpose of kinetic theory is to model the time evolution of a
collection of particles. The particles may be entirely different
objects depending on the physical situation. For instance, the
particles are atoms and molecules in a neutral gas or electrons and
ions in a plasma. In astrophysics the particles are stars, galaxies
or even clusters of galaxies. Mathematical models of particle systems
are most frequently described by kinetic or fluid equations.
A characteristic feature of kinetic theory is that its models are statistical
and the particle
systems are described by density functions $f=f(t,x,p)$, which
represent the density of particles with given spacetime position
$(t,x)\in\mathbb{R}\times\mathbb{R}^3$ and momentum $p\in
\mathbb{R}^{3}$. A density function contains a wealth of
information, and macroscopic quantities are easily calculated from
this function. In a fluid model the quantities that describe the
system do not depend on the momentum $p$ but only on the spacetime
point $(t,x)$. A choice of model is usually made with regard to the
physical properties of interest for the system or with regard to
numerical considerations. It should be mentioned that a too naive
fluid model may give rise to shell-crossing singularities, which
are unphysical. In a kinetic description such phenomena are ruled out.

The time evolution of the system is determined by the interactions
between the particles, which depend on the physical situation. For
instance, the driving mechanism for the time evolution of a neutral
gas is the collision between particles (the Boltzmann equation).
For a plasma the interaction is through the electromagnetic field
produced by the charges (the Vlasov--Maxwell system), and in
astrophysics the interaction is gravitational (the Vlasov--Poisson
system and the Einstein--Vlasov system). Of course, combinations of
interaction processes are also considered but in many situations one
of them is strongly dominating and the weaker processes are neglected.


\subsection{The relativistic Boltzmann equation}

Consider a collection of neutral particles in Minkowski spacetime. Let
the signature of the metric be $(-,+,+,+)$. In this section we assume
that all the particles have rest mass $m=1$, and we normalize the
speed of light $c$ to one. We point out that in
Section~\ref{sec:einstein-vlasov-system} on the Einstein--Vlasov
system, the dependence on the rest mass and the speed of light will be
included in the formulation of the system. The
four-momentum of a particle is denoted by $p^{a}$, $a=0,1,2,3$. Since
all particles have equal rest mass, the four-momentum for each
particle is restricted to the mass shell, $p^{a}p_{a}=-m^{2}=-1$. Thus,
by denoting the three-momentum by $p\in\mathbb{R}^{3}$, $p^{a}$ may
be written $p^{a}=(p^0,p)$, where
$p^{0}=\sqrt{1+|p|^{2}}$ is the energy of a particle with
three-momentum $p$, and $|p|$ is the usual Euclidean length
of $p$. The relativistic velocity of a particle with
momentum $p$ is denoted by $\hat{p}$ and is given by
\begin{equation}
  \hat{p}=\frac{p}{\sqrt{1+|p|^{2}}}.
  \label{velo}
\end{equation}
Note that $|\hat{p}|<1=c$. The relativistic Boltzmann equation models
the spacetime behavior of the one-particle distribution function
$f=f(t,x,p)$, and it has the form
\begin{equation}
  \left( \partial_{t} + \frac{p}{p^{0}} \cdot \nabla_{x} \right) f = Q(f,f),
  \label{rbe}
\end{equation}
where the relativistic collision operator $Q(f,g)$ is defined by
\begin{equation}
  Q(f,g) = \int_{\mathbb{R}^{3}} \int_{\mathbb{S}^{2}} k(p,q,\omega)
  [f(p+a(p,q,\omega)\omega) \, g(q-a(p,q,\omega)\omega)-f(p)g(q)] \,
  d\omega \, dp.
  \label{rbeQ}
\end{equation}
Note that $g=f$ in Equation~(\ref{rbe}). Here $d\omega$ is the element of
surface area on $\mathbb{S}^{2}$ and $k(p,q,\omega)$ is the scattering
kernel, which depends on the differential cross-section in the
interaction process. We refer to~\cite{Srm1u2}, \cite{CeKru2} and \cite{GLW}
for examples of differential cross-sections in the relativistic case.
The function $a(p,q,\omega)$ results from the
collision mechanics. If two particles, with momentum $p$ and $q$
respectively, collide elastically with scattering
angle $\omega\in\mathbb{S}^{2}$, their momenta will change, i.e.,
$p\rightarrow p'$ and $q\rightarrow q'$. The relation between $p,q$
and $p',q'$ is given by
\begin{equation}
  p'=p+a(p,q,\omega)\omega,
  \qquad
  q'=q-a(p,q,\omega)\omega,
\end{equation}
where
\begin{equation}
  a(p,q,\omega)=\frac{2(p^{0}+q^{0})p^{0}
    q^{0}(\omega\cdot(\hat{q}-\hat{p}))}
  {(p^{0}+q^{0})^{2}-(\omega\cdot(p+q))^{2}}.
\end{equation}
This relation is a consequence of four-momentum conservation,
\begin{displaymath}
  p^{a}+q^{a}=p^{a'}+q^{a'},
\end{displaymath}
or equivalently
\begin{eqnarray}
  p^{0}+q^{0}&=&p^{0'}+q^{0'}\!\!,
  \\
  p+q&=&p'+q'.
\end{eqnarray}
These are the conservation equations for relativistic particle
dynamics. In the classical case the corresponding conservation
equations read
\begin{eqnarray}
  |p|^{2}+|q|^{2}&=&|p'|^{2}+|q'|^{2},
  \label{cocl1}
  \\
  p+q&=&p'+q'
  \label{cocl2}.
\end{eqnarray}
The function $a(p,q,\omega)$ gives the distance between $p$ and $p'$ ($q$
and $q'$) in momentum space, and the analogue function in the non-relativistic,
Newtonian, classical case has the form
\begin{equation}
  a_\mathrm{cl}(p,q,\omega)=\omega\cdot (q-p).
\end{equation}
By inserting $a_\mathrm{cl}$ in place of $a$ in Equation~(\ref{rbeQ})
we obtain the classical Boltzmann collision operator (disregarding the
scattering kernel, which is also different). We point out that there are
other representations of the collision operator~(\ref{rbeQ}), cf.~\cite{Srm3u2}.

In~\cite{C1} and \cite{Srm1u2} classical
solutions to the relativistic Boltzmann equations are studied as
$c\to\infty$, and it is proven that the limit as $c\to\infty$ of these
solutions satisfies the classical Boltzmann equation. The former work is more general 
since general initial data is considered, whereas the latter is concerned with data near 
vacuum. The latter result is stronger in the sense that the limit, as $c\to\infty,$ is 
shown to be uniform in time. 

The main result concerning the existence of solutions to the classical
Boltzmann equation is a theorem by DiPerna and Lions~\cite{DL1} that
proves existence, but not uniqueness, of renormalized solutions.
An analogous result holds in the relativistic case, as was shown
by Dudy\'{n}ski and Ekiel-Je{\.z}ewska~\cite{DE}, cf.\ also~\cite{Ju2}.
Regarding classical solutions,
Illner and Shinbrot~\cite{IlS} have shown global existence of solutions
to the non-relativistic Boltzmann equation for initial data close to vacuum.
Glassey showed global existence for data near vacuum in the relativistic case 
in a technical work~\cite{Gu2}. He only requires decay and integrability conditions 
on the differential cross-section, although these are not fully
satisfactory from a physics point of view. 
By imposing more restrictive cut-off assumptions 
on the differential cross-section, Strain~\cite{Srm1u2} gives a different proof, 
which is more related to the proof
in the non-relativistic case~\cite{IlS} than~\cite{Gu2} is. 
For the homogeneous
relativistic Boltzmann equation, global existence for small initial data
has been shown in~\cite{NT} under the assumption of a bounded differential
cross-section. For initial data close to equilibrium, global existence of
classical solutions has been proven by Glassey and Strauss~\cite{GSt3} using assumptions
on the differential cross-section, which fall into the regime ``hard potentials'',
whereas Strain~\cite{Srm2u2} has shown existence in the case of soft potentials.
Rates of the convergence to equilibrium are given in both~\cite{GSt3} and \cite{Srm2u2}.
In the non-relativistic case, we refer to~\cite{U,Sh,NI} for analogous results.

The collision operator $Q(f,g)$ may be written in an obvious way as
\begin{displaymath}
  Q(f,g)=Q^{+}(f,g)-Q^{-}(f,g),
\end{displaymath}
where $Q^{+}$ and $Q^{-}$ are called the gain and loss term,
respectively. If the loss term is deleted the gain-term-only Boltzmann
equation is obtained. It is interesting to note that the methods of
proof for the small data results mentioned above concentrate on
gain-term-only equations, and once that is solved it is easy to
include the loss term. In~\cite{ACI} it is shown that the
gain-term-only classical and relativistic Boltzmann equations blow up
for initial data not restricted to a small neighborhood of trivial
data. Thus, if a global existence proof of classical solutions for
unrestricted data will be given, it will necessarily use the full collision operator.

The gain term has a nice regularizing property in the momentum
variable. In~\cite{An1} it is proven that given $f\in
L^{2}(\mathbb{R}^{3})$ and $g\in L^{1}(\mathbb{R}^{3})$ with
$f,g \geq 0$, then
\begin{equation}
  \| Q^{+}(f,g)\|_{H^{1}(\mathbb{R}_{p}^{3})}\leq C\| f\|_{L^{2}
  (\mathbb{R}_{p}^{3})}\|g\|_{L^{1}(\mathbb{R}_{p}^{3})},
  \label{Q}
\end{equation}
under some technical requirements on the scattering kernel. Here
$H^{s}$ is the usual Sobolev space. This regularizing result was first
proven by Lions~\cite{Li} in the classical situation. The proof
relies on the theory of Fourier integral operators and on the method
of stationary phase, and requires a careful analysis of the collision
geometry, which is very different in the relativistic case. Simplified
proofs in the classical and relativistic case are given in~\cite{Wen1,Wen2}.

The regularizing theorem has many applications. An important
application is to prove that solutions tend to equilibrium for large
times. More precisely, Lions used the regularizing theorem to prove
that solutions to the classical Boltzmann equation, with periodic
boundary conditions, converge in $L^{1}$ to a global Maxwellian,
\begin{displaymath}
  M=e^{-\alpha |p|^{2}+\beta\cdot p+\gamma}
  \qquad
  \text{ with }
  \alpha,\gamma\in R,
  \quad
  \alpha>0,
  \quad
  \beta\in \mathbb{R}^{3},
\end{displaymath}
as time goes to infinity. This result was first obtained by
Arkeryd~\cite{Ar} by using non-standard analysis. It should be pointed
out that the convergence takes place through a sequence of times
tending to infinity and it is not known whether the limit is unique or
depends on the sequence. In the relativistic situation, the analogous
question of convergence to a relativistic Maxwellian, or a J\"{u}ttner
equilibrium solution,
\begin{displaymath}
  J=e^{-\alpha\sqrt{1+|p|^{2}}+\beta\cdot p+\gamma},
  \qquad
  \alpha,\beta, \text{ and } \gamma \text{ as above, with } \alpha>|\beta|,
\end{displaymath}
was studied by Glassey and Strauss~\cite{GSt3, GSt4}. In the
periodic case, they proved convergence in a variety of function spaces
for initial data close to a J\"{u}ttner solution. Having obtained the
regularizing theorem for the relativistic gain term, it is a
straightforward task to follow the method of Lions and prove
convergence to a global J\"{u}ttner solution for arbitrary initial data
(satisfying the natural bounds of finite energy and entropy), which are
periodic in the space variables, cf.~\cite{An1}. We also mention
that in the non-relativistic case Desvillettes and Villani~\cite{DVu2} have studied
the convergence rate to equilibrium in detail.
A similar study in the relativistic case has not yet been achieved.

For more information on the relativistic Boltzmann equation on
Minkowski space we refer to~\cite{CeKru2, GLW, Sy, Gl} and in the
non-relativistic case we refer to~\cite{V,Gl,CIP}.


\subsection{The Vlasov--Maxwell and Vlasov--Poisson systems}

Let us consider a collision-less plasma, which is a collection of
particles for which  collisions are relatively rare and the
interaction is through their charges. For simplicity we assume that the plasma
consists of one type of particle, although the results below
hold for plasmas with several particle species. The particle rest mass
and the particle charge are normalized to one. In the kinetic
framework, the most general set of equations for modeling a
collision-less plasma is the relativistic Vlasov--Maxwell system:
\begin{eqnarray}
  \partial_{t}f+\hat{v}\cdot\nabla_{x}f+(E(t,x)+\hat{v}\times
  B(t,x))\cdot\nabla_{v}f&=&0
  \label{rvm}
  \\ [0.5 em]
  \partial_{t}E+j=c\nabla\times B,
  \hphantom{-}\qquad
  \nabla\cdot E&=&\rho,
  \label{maxwell}
  \\
  \partial_{t}B=-c\nabla\times E,
  \qquad
  \nabla\cdot B&=&0.
  \label{maxwell_2}
\end{eqnarray}
The notation follows the one already introduced with the exception
that the momenta are now denoted by $v$ instead of $p$. This has
become a standard notation in this field. $E$ and $B$ are the electric
and magnetic fields, and $\hat{v}$ is the relativistic velocity,
\begin{equation}
  \hat{v}=\frac{v}{\sqrt{1+|v|^2/c^2}},
  \label{relveldef}
\end{equation}
where $c$ is the speed of light. The charge density $\rho$ and current
$j$ are given by
\begin{equation}
  \rho=\int_{\mathbb{R}^{3}} fdv,
  \qquad
  j=\int_{\mathbb{R}^{3}} \hat{v}fdv.
\end{equation}
Equation~(\ref{rvm}) is the relativistic Vlasov equation and
Equations~(\ref{maxwell}, \ref{maxwell_2}) are the Maxwell equations.

A special case in three dimensions is obtained by considering
spherically-symmetric initial data. For such data it can be shown that
the solution will also be spherically symmetric, and that the magnetic
field has to be constant. The Maxwell equation
$\nabla\times E=-\partial_tB$ then implies that the electric field is
the gradient of a potential $\phi$. Hence, in the spherically-symmetric case the relativistic Vlasov--Maxwell system takes the form
\begin{eqnarray}
  \partial_{t}f+\hat{v}\cdot\nabla_{x}f+\beta E(t,x) \cdot\nabla_{v}f&=&0,
  \label{rv}
  \\
  E=\nabla\phi,
  \qquad
  \Delta\phi&=&\rho.
  \label{rpoisson}
\end{eqnarray}
Here $\beta=1$, and the constant magnetic field has been set to zero,
since a constant field has no significance in this discussion. This
system makes sense for any initial data, without symmetry constraints,
and is called the relativistic Vlasov--Poisson system. Another
special case of interest is the classical limit, obtained by letting
$c\rightarrow\infty$ in Equations~(\ref{rvm}, \ref{maxwell},
\ref{maxwell_2}), yielding:
\begin{eqnarray}
  \partial_{t}f+v\cdot\nabla_{x}f+\beta E(t,x) \cdot\nabla_{v}f=0,
  \label{v}
  \\
  E=\nabla\phi,
  \qquad
  \Delta\phi&=&\rho,
  \label{poisson}
\end{eqnarray}
where $\beta=1$. We refer to Schaeffer~\cite{Sc1} for a rigorous derivation of
this result. This is the Vlasov--Poisson system,
and $\beta=1$ corresponds to repulsive forces (the plasma
case). Taking $\beta=-1$ means attractive forces and the
Vlasov--Poisson system is then a model for a Newtonian
self-gravitating system.

One of the fundamental problems in kinetic theory is to find out
whether or not spontaneous shock formations will develop in a
collision-less gas, i.e., whether solutions to any of the equations
above will remain smooth for all time, given smooth initial data.

If the initial data are small this problem has an affirmative solution
in all cases considered above~\cite{GSc1, GSt2, BD, BDH}. For
initial data unrestricted in size the picture is more involved. In
order to obtain smooth solutions globally in time, the main issue is
to control the support of the momenta
\begin{equation}
  Q(t):=\sup \{|v|:\exists (s,x)\in [0,t]\times \mathbb{R}^3
  \text{ such that }
  f(s,x,v)\not= 0\},
  \label{Qnew}
\end{equation}
i.e., to bound $Q(t)$ by a continuous function so that $Q(t)$ will not
blow up in finite time. That such a control is sufficient for
obtaining global existence of smooth solutions follows from well-known
results in the different cases, cf.~\cite{GSt1, KS, BGP, Ho1, B, GSc1}. For
the full three-dimensional relativistic Vlasov--Maxwell system, the
problem of establishing whether or not solutions will remain
smooth for all time is open.
A different sufficient criterion for global
existence in this case is given by Pallard in~\cite{P1}, and he also shows a new
bound for the electromagnetic field in terms of $Q(t)$ in~\cite{P2}.
In two space and three momentum dimensions, Glassey and Schaeffer~\cite{GSc2,GSc3}
have shown that $Q(t)$ can be controlled for
the relativistic Vlasov--Maxwell system, which thus yields
global existence of smooth solutions in that case.

The relativistic and non-relativistic Vlasov--Poisson equations are
very similar in form. In particular, the equation for the field is
identical in the two cases. However, the mathematical results
concerning the two systems are very different. In the non-relativistic
case, Batt~\cite{B} gave an affirmative solution in 1977 in the case of
spherically-symmetric data. Pfaffelmoser~\cite{Pf}
was the first one to give a proof for general
smooth data. A simplified version of the proof is given by Schaeffer in~\cite{Sc2}.
Pfaffelmoser obtained the bound
\begin{displaymath}
  Q(t)\leq C(1+t)^{(51+\delta)/11},
\end{displaymath}
where $\delta>0$ can be taken as arbitrarily small. This bound was
later improved by different authors. The sharpest bound valid for
$\beta=1$ and $\beta=-1$ has been given by Horst~\cite{Ho2} and reads
\begin{displaymath}
  Q(t)\leq C(1+t)\log(2+t).
\end{displaymath}
In the case of repulsive forces ($\beta=1$) Rein~\cite{Rn1} has found
a better estimate by using a new identity for the Vlasov--Poisson
system, discovered independently by Illner and Rein~\cite{IlR} and
by Perthame~\cite{Pe}. Rein's estimate reads
\begin{displaymath}
  Q(t)\leq C(1+t)^{2/3}.
\end{displaymath}
Independently, and at about the same time as Pfaffelmoser gave his proof,
Lions and Perthame~\cite{LP} used a different method for proving
global existence. Their method is more
generally applicable, and the two studies~\cite{An2} and~\cite{KR} are examples
of problems in related systems, where their method has been successful.
On the other hand, their method does not
give such strong growth estimates on $Q(t)$ as described above. For
the relativistic Vlasov--Poisson equation, Glassey and
Schaeffer~\cite{GSc1} showed in the case $\beta=1$ that if the data
are spherically symmetric, $Q(t)$ can be controlled, which is
analogous to the result by Batt mentioned above. Also in the case of
cylindrical symmetry they are able to control $Q(t)$;
see~\cite{GSc4}. If $\beta=-1$ it was shown
in~\cite{GSc1} that blow-up occurs in finite time for spherically-symmetric data with negative total energy. More recently, Lemou et al.~\cite{LMRu2}
have investigated the structure of the blow-up solution. They
show that the blow-up is determined by the self-similar solution of
the ultra-relativistic gravitational Vlasov--Poisson system.
It should be pointed out that the relativistic Vlasov--Poisson system is
unphysical since it lacks the Lorentz invariance; it
is a hybrid of a classical Galilei invariant field equation and a
relativistic transport equation~(\ref{rv}), cf.~\cite{An3}.
In particular, in the case $\beta=-1$,
it is not a special case of the Einstein--Vlasov system.
Only for spherically-symmetric data, in the case $\beta=1$, is the equation a
fundamental physical equation.
The results mentioned above all concern classical solutions. The situation
for weak solutions is different, in particular the existence of weak
solutions to the relativistic Vlasov--Maxwell system is known~\cite{DL2, Rn7}.

We also mention that models, which take into account both collisions and
the electric and magnetic fields generated by the particles have been investigated.
Classical solutions near a Maxwellian for the Vlasov--Maxwell--Boltzmann system are
constructed by Guo in~\cite{Guou2}. A similar result for the Vlasov--Maxwell--Landau
system near a J\"{u}ttner solution is shown by Guo and Strain in~\cite{GSu2}.

We refer to the book by Glassey~\cite{Gl} and the review article by Rein~\cite{Rn1u2} for
more information on
the relativistic Vlasov--Maxwell system and the Vlasov--Poisson system.

\subsection{The Nordstr\"{o}m--Vlasov system}

Before turning to the main theme of this review, i.e., the Einstein--Vlasov
system, we briefly review the results on the Nordstr\"{o}m--Vlasov system. Nordstr\"{o}m
gravity~\cite{Nor} is an alternative theory of gravity introduced in 1913.
By coupling this model to a kinetic description of matter
the Nordstr\"{o}m--Vlasov system results. In Nordstr\"{o}m gravity
the scalar field $\phi$ describes the gravitational field in the sense given below.
The Nordstr\"{o}m--Vlasov system reads
\begin{eqnarray}
  &&\partial_t^2\phi-\bigtriangleup_x\phi=-e^{4\phi}
  \int_{\mathbb{R}^3}\frac{\mathfrak{f}\,dp}{\sqrt{1+|p|^{2}}},
  \label{wave1}
  \\
  &&\partial_{t}\mathfrak{f} + \widehat{p}\cdot\nabla_x\mathfrak{f} -
  \left[\left(\partial_{t}\phi + \widehat{p}\cdot\nabla_x\phi \right) p +
    (1+|p|^{2})^{-1/2}\nabla_x\phi\right]\cdot\nabla_p\mathfrak{f}=0.
  \label{vlasov01}
\end{eqnarray}
Here
\begin{displaymath}
  \widehat{p} =\frac{p}{\sqrt{1+|p|^2}},
\end{displaymath}
denotes the relativistic velocity of a particle with momentum $p$.
The mass of each particle,
the gravitational constant, and the speed of light are
all normalized to one.
A solution $(\mathfrak{f},\phi)$ of this system
is interpreted as follows. The spacetime is a
Lorentzian manifold with a conformally-flat metric
\begin{displaymath}
g_{\mu\nu}=e^{2\phi} \diag (-1,1,1,1).
\end{displaymath}
The particle distribution $f$ defined on the mass shell
in this metric is given by
\begin{equation}
  f(t,x,p)=\mathfrak{f}(t,x,e^\phi p).
  \label{fph}
\end{equation}
%
The first mathematical study of this system was initiated by Calogero in~\cite{C2},
where the existence of static solutions is established.
The stability of the static solutions was then investigated in~\cite{CSSu2}.
Although the Nordstr\"{o}m--Vlasov model of gravity does not describe physics correctly,
the system approaches the Vlasov--Poisson system in the classical limit. Indeed, it
is shown in~\cite{CL} that solutions of the Nordstr\"{o}m--Vlasov
system tend to solutions of the Vlasov--Poisson system as the speed of
light goes to infinity.

The Cauchy problem was studied by several authors~\cite{CR1,CR2,ACR,Le3,P3} and the question
of global existence of classical solutions for general initial data was open for some time. The problem was
given an affirmative solution in 2006 by Calogero~\cite{Cal3u2}. Another interesting result 
for the Nordstr\"{o}m--Vlasov system is given in~\cite{BKRRu2}, where a radiation formula,
similar to the dipole formula in electrodynamics, is rigorously derived.

\newpage

\section{The Einstein--Vlasov System}
\label{sec:einstein-vlasov-system}

In this section we consider a self-gravitating collision-less gas in the framework
of general relativity and we present the Einstein--Vlasov system. It is most often
the case in the mathematics literature that the speed of light $c$ and the gravitational
constant $G$ are normalized to one, but we keep these constants in the formulas
in this section
since in some problems they do play an important role. However, in most of the problems
discussed in the forthcoming sections these constants will be normalized to one.

Let $M$ be a four-dimensional manifold and let $g_{ab}$ be a metric
with Lorentz signature $(-,+,+,+)$ so that $(M,g_{ab})$ is a
spacetime.
The metric is assumed to be time-orientable so that there is a distinction between
future and past directed vectors.

The possible values of the four-momentum $p^{a}$ of a particle with rest mass $m$ belong
to the mass shell $P_m\subset TM$, defined by
\begin{equation}
\label{massshell}
P_m:=\{(x^{a},p^{a})\in\, TM\,:\,g_{ab}(x^a)p^{a}p^{b}=-m^2c^2,\;p^{a}
\text{ is future directed} \}.
\end{equation}
Hence, if $m>0$, $P_m(x^{a})$ is the set of all future-directed
time-like vectors with length $cm$, and if $m=0$ it is the set of all
future-directed null vectors. On $P_m$ we take $(x^a,p^j),\,a=0,1,2,3$
and $j=1,2,3$ (letters in the beginning of the alphabet always take
values $0,1,2,3$ and letters in the middle take $1,2,3$) as local
coordinates, and $p^0$ is expressed in terms of $p^j$ and the metric
in view of Equation~(\ref{massshell}). Thus, the density function $f_m$ is
a non-negative function on $P_m$. Below we drop the index $m$ on
$f_m$ and simply write $f$.

Since we are considering a collisionless gas, the particles follow the geodesics in
spacetime. The geodesics are projections onto spacetime of the curves in $P_m$ defined
in local coordinates by
\begin{eqnarray}
\frac{dx^{a}}{ds}&=&p^{a},\nonumber\\
\frac{dp^{j}}{ds}&=&-\Gamma^{j}_{bc}p^{b}p^{c}.\nonumber
\end{eqnarray}
Here $\Gamma_{bc}^{a}$ are the Christoffel symbols. Along a geodesic the density
function $f=f(x^a,p^j)$ is invariant so that
\[
\frac{d}{ds}f(x^a(s),p^j(s))=0,
\]
which implies that
\begin{equation}
  p^{a}\frac{\partial f}{\partial x^a}-\Gamma^{j}_{ab}p^ap^b\frac{\partial f}{\partial p^j}=0.
  \label{vlasov1}
\end{equation}
This is accordingly the Vlasov equation. We point out that sometimes the density
function is considered as a function on the entire tangent bundle $TM$ rather than on
the mass shell $P_m\subset TM$. The Vlasov equation for $f=f(x^a,p^a)$ then takes the form
%
\begin{equation}
  p^{a}\frac{\partial f}{\partial x^a}-\Gamma^{a}_{bc}p^bp^c\frac{\partial f}{\partial p^a}=0.
  \label{vlasov2}
\end{equation}
This equation follows from~(\ref{vlasov1}) if we take the mass shell condition
$g_{ab}p^ap^b=-m^2c^2$ into account. Indeed, by abuse of notation, we have
\begin{eqnarray}
  \frac{\partial f}{\partial x^a}&=&\frac{\partial f}{\partial x^a}+\frac{\partial f}{\partial p^0}\frac{\partial p^0}{\partial x^a},\nonumber\\
  \frac{\partial f}{\partial p^j}&=&\frac{\partial f}{\partial p^j}+\frac{\partial f}{\partial p^0}\frac{\partial p^0}{\partial p^j}.\nonumber
\end{eqnarray}
Here $f$ is considered as a function on $P_m$ in the left-hand side,
and on $TM$ in the right-hand side.
From the mass shell condition $g_{ab}p^ap^b=-m^2c^2$ we derive
\begin{eqnarray}
  \frac{\partial p^0}{\partial x^a}&=&-\frac{1}{p_0}p^bp_c\Gamma_{ab}^c,\nonumber\\
  \frac{\partial p^0}{\partial p^j}&=&-\frac{p_j}{p_0}.\nonumber
\end{eqnarray}
Inserting these relations into (\ref{vlasov1}) we obtain (\ref{vlasov2}). If we
let $t=x^0$, and divide the Vlasov equation (\ref{vlasov1}) by $p^0$
we obtain the most common form in the literature of the Vlasov equation
\begin{equation}
  \frac{\partial f}{\partial t}+\frac{p^{j}}{p^0}\frac{\partial f}{\partial x^j}-\frac{1}{p^0}
  \Gamma^{j}_{ab}p^ap^b\frac{\partial f}{\partial p^j}=0.
  \label{vlasovgamma}
\end{equation}
In a fixed spacetime the Vlasov equation (\ref{vlasovgamma}) is a linear hyperbolic
equation for $f$ and we can solve it by solving the characteristic
system,
\begin{eqnarray}
  \frac{dX^i}{ds}&=&\frac{P^i}{P^0},
  \label{char1}
  \\
  \frac{dP^i}{ds}&=&-\Gamma^i_{ab}\frac{P^aP^b}{P^0}.
  \label{char2}
\end{eqnarray}
In terms of initial data $f_0$ the solution of the Vlasov equation can
be written as
\begin{equation}
  f(x^a,p^i)=f_0(X^i(0,x^a,p^i),P^i(0,x^a,p^i)),
  \label{solution}
\end{equation}
where $X^i(s,x^a,p^i)$ and $P^i(s,x^a,p^i)$ solve
Equations~(\ref{char1}, \ref{char2}), and where
$$
X^i(t,x^a,p^i)=x^i
\text{ and }
P^i(t,x^a,p^i)=p^i.
$$

In order to write down the Einstein--Vlasov system we need to know
the energy-momentum tensor $T_{ab}^m$ in terms of $f$ and $g_{ab}$. We define
\begin{equation}
\label{enermomts}
  T_{ab}^m=c\sqrt{|g_{ab}|}\int_{\mathbb{R}^3}f\,p_ap_b\frac{dp^1dp^2dp^3}{-p_0},
\end{equation}
where, as usual, $p_a=g_{ab}p^b$, and $|g_{ab}|$ denotes the absolute value of
the determinant of $g_{ab}$. We remark that the measure
\[
\mu:=\frac{\sqrt{|g_{ab}|}}{-p_0}dp^1dp^2dp^3,
\]
is the induced metric of the submanifold $P_m(x^a)\subset T_{x^a}M$,
and that $\mu$ is invariant under Lorentz transformations of the
tangent space, and it is often the case in the literature that
$T_{ab}^m$ is written as
\begin{displaymath}
  T_{ab}^m=c\int_{P_m(x^a)}f\,p_ap_b\,\mu.
\end{displaymath}

Let us now consider a collisionless gas consisting of particles with different
rest masses 
\\$m_1,m_2,\dots,m_N$, described by $N$ density functions $f_{m_j},\,j=1,\dots,N$. 
Then the Vlasov equations for the different density functions
$f_{m_j}$, together with the Einstein equations,
\begin{displaymath}
\label{eineqs}
R_{ab}-\frac{1}{2}Rg_{ab}+\Lambda g_{ab}=\frac{8\pi\,G}{c^4} \sum_{k=1}^{N}T_{ab}^{m_{k}},
\end{displaymath}
form the Einstein--Vlasov system for the collision-less gas.
Here $R_{ab}$ is the Ricci tensor,
$R$ is the scalar curvature and $\Lambda$ is the cosmological constant.

Henceforth, we always assume that there is only one species
of particles in the gas and we write $T_{ab}$ for its energy
momentum tensor. Moreover, in what follows, we normalize the rest mass $m$
of the particles, the speed of
light $c$, and the gravitational constant $G$, to one, if not otherwise explicitly
stated that this is not the case.

Let us now investigate the features of the energy momentum tensor for Vlasov matter.
We define the particle current density
\begin{displaymath}
  N^a=-\int_{\mathbb{R}^{3}}f\,p^a\sqrt{|g_{ab}|}\frac{dp^1\,dp^2\,dp^3}{p_0}.
\end{displaymath}
Using normal coordinates based at a given point and assuming that $f$
is compactly supported, it is not hard to see that $T_{ab}$ is
divergence-free, which is a necessary compatibility condition since
the left-hand side of~(\ref{eineqs}) is divergence-free by the Bianchi identities.
A computation in normal coordinates also shows that $N^a$ is divergence-free, which
expresses the fact that the number of particles is conserved. The
definitions of $T_{ab}$ and $N^a$ immediately give us a number of
inequalities. If $V^a$ is a future-directed time-like or null vector
then we have $N_aV^a\leq 0$ with equality if and only if $f=0$ at the
given point. Hence, $N^a$ is always future-directed time-like, if
there are particles at that point. Moreover, if $V^a$ and $W^a$ are
future-directed time-like vectors then $T_{ab}V^aW^b\geq 0$, which is
the dominant energy condition. This also implies that the weak energy
condition holds. If $X^a$ is a space-like vector, then
$T_{ab}X^aX^b\geq 0$. This is called the non-negative pressure
condition, and it implies that the strong energy condition holds as
well. That the energy conditions hold for Vlasov matter is one reason
that the Vlasov equation defines a well-behaved matter model in general
relativity. Another reason is the well-posedness theorem by
Choquet-Bruhat~\cite{Ct} for the Einstein--Vlasov system that we state
below. Before stating that theorem we first discuss the conditions
imposed on the initial data.

The initial data in the Cauchy problem for the Einstein--Vlasov system consist
of a 3-dimensional manifold $S$, a Riemannian metric $g_{ij}$ on $S$,
a symmetric tensor $k_{ij}$ on $S$, and a non-negative scalar function
$f_0$ on the tangent bundle $TS$ of $S$.

The relationship between a given initial data set
$(g_{ij},k_{ij})$ on $S$ and the metric
$g_{ab}$ on the spacetime manifold is, that there exists an embedding
$\psi$ of $S$ into the spacetime such that the induced metric and
second fundamental form of $\psi(S)$ coincide with the result of
transporting $(g_{ij},k_{ij})$ with $\psi$. For the relation of the
distribution functions $f$ and $f_0$ we have to note that $f$ is
defined on the mass shell. The initial condition imposed is that the
restriction of $f$ to the part of the mass shell over $\psi(S)$ should
be equal to $f_0\circ (\psi^{-1},d(\psi)^{-1})\circ\phi$, where $\phi$
sends each point of the mass shell over $\psi(S)$ to its orthogonal
projection onto the tangent space to $\psi(S)$. An initial data set
for the Einstein--Vlasov system must satisfy the constraint equations,
which read
\begin{eqnarray}
  R-k_{ij}k^{ij}+(\tr k)^2&=&16\pi\rho,
  \label{constr1}
  \\
  \nabla_{i}k^{i}_{l}-\nabla_{l}(\tr k)&=&8\pi j_l.
  \label{constr2}
\end{eqnarray}
Here $\rho=T_{ab}n^an^b$ and $j^a=-h^{ab}T_{bc}n^c$, where $n^a$ is
the future directed unit normal vector to the initial hypersurface, and
$h^{ab}=g^{ab}+n^an^b$ is the orthogonal projection onto the tangent
space to the initial hypersurface. In terms of $f_0$ we can express
$\rho$ and $j^l$ by ($j^a$ satisfies $n_aj^a=0$, so it can naturally be
identified with a vector intrinsic to $S$)
\begin{eqnarray*}
  \rho&=&\int_{\mathbb{R}^{3}}f_0\,p^ap_a\,\sqrt{|g_{ij}|}\frac{dp^1\,dp^2\,dp^3}
  {1+p_jp^j},
  \\
  j_l&=&\int_{\mathbb{R}^{3}}f_0\,p_l\sqrt{|g_{ij}|}\, dp^1\,dp^2\,dp^3.
\end{eqnarray*}
%
We can now state the local existence theorem by
Choquet-Bruhat~\cite{Ct}, for the Einstein--Vlasov system.

\begin{theorem}
  Let $S$ be a 3-dimensional manifold, $g_{ij}$ a smooth Riemannian
  metric on $S$, $k_{ij}$ a smooth symmetric tensor on $S$ and $f_0$ a
  smooth non-negative function of compact support on the tangent bundle
  $TS$ of $S$. Suppose that these objects satisfy the constraint
  equations~(\ref{constr1}, \ref{constr2}). Then there exists a
  smooth spacetime $(M,g_{ab})$, a smooth distribution function $f$ on
  the mass shell of this spacetime, and a smooth embedding $\psi$ of $S$
  into $M$, which induces the given initial data on $S$ such that
  $g_{ab}$ and $f$ satisfy the Einstein--Vlasov system and $\psi(S)$ is
  a Cauchy surface. Moreover, given any other spacetime $(M',g'_{ab})$,
  distribution function $f'$ and embedding $\psi'$ satisfying these
  conditions, there exists a diffeomorphism $\chi$ from an open
  neighborhood of $\psi(S)$ in $M$ to an open neighborhood of
  $\psi'(S)$ in $M'$, which satisfies $\chi\circ\psi=\psi'$ and carries
  $g_{ab}$ and $f$ to $g'_{ab}$ and $f'$, respectively.
\end{theorem}

The above formulation is in the case of smooth initial data; for information 
on the regularity needed on the initial data we refer to~\cite{Ct} and~\cite{Mu1u2}.
In this context we also mention that local existence has been proven
for the Yang--Mills--Vlasov system in~\cite{CN}, and that this problem
for the Einstein--Maxwell--Boltzmann system is treated
in~\cite{BC}. However, this result is not complete, as the non-negativity
of $f$ is left unanswered. Also, the hypotheses on the scattering
kernel in this work leave some room for further investigation. The
local existence problem for physically reasonable assumptions on the scattering
kernel does not seem well understood in the context of the
Einstein--Boltzmann system, and a careful study of this problem would be
desirable. The mathematical study of the Einstein--Boltzmann system has
been very sparse in the last few decades, although there has been some activity in recent years.
Since most questions on the global properties are completely open let us only very
briefly mention some of these works.
Mucha~\cite{Mu2u2} has improved the regularity assumptions on the initial data assumed 
in~\cite{BC}.
Global existence for the homogeneous Einstein--Boltzmann system in Robertson--Walker
spacetimes is proven in~\cite{NTau2}, and a generalization to Bianchi type I
symmetry is established in~\cite{NDu2}. 

In the following sections we present results on the global properties
of solutions of the Einstein--Vlasov system, which have been obtained during the last
two decades.

Before ending this section we mention a few other sources for more background
on the Einstein--Vlasov system, cf.~\cite{Rl6,Rl2u2,E,St}.


\newpage

\section{The Asymptotically-Flat Cauchy Problem: Spherically-Symmetric Solutions}

In this section, we discuss results on global existence and on the asymptotic structure
of solutions of the Cauchy problem in the asymptotically-flat case. 

In general relativity two classes of initial data are
distinguished in the study of the Cauchy problem: asymptotically-flat initial
data and cosmological initial data. The former type of data describes
an isolated body. The initial hypersurface is
topologically $\mathbb{R}^{3}$ and appropriate fall-off conditions
are imposed to ensure that far away from the body spacetime is approximately flat.
Spacetimes, which possess a compact Cauchy hypersurface, are called
cosmological spacetimes, and data are accordingly given on a compact
3-manifold. In this case, the whole universe is modeled rather than
an isolated body.

The symmetry classes that admit
asymptotic flatness are few. The important ones are spherically
symmetric and axially symmetric spacetimes. 
One can also consider a case, which is unphysical, in which spacetime is asymptotically flat except in one
direction, namely cylindrically-symmetric spacetimes, cf.~\cite{F}, where the Cauchy problem is studied. 
The majority of the work so far has been devoted to
the spherically-symmetric case but recently a result on static axisymmetric solutions has been obtained. 

In contrast to the asymptotically-flat case,
cosmological spacetimes admit a large number of symmetry classes. This
provides the possibility to study many special cases for which
the difficulties of the full Einstein equations are
reduced. The Cauchy problem in the cosmological case is reviewed
in Section~\ref{sec:cosmological-cauchy-problem}.

The following subsections concern studies of the spherically-symmetric Einstein--Vlasov
system. The main goal of these studies is to provide an answer to the weak and strong
cosmic censorship conjectures, cf.~\cite{Wa,Cu1u2} for formulations of the conjectures.

\subsection{Set up and choice of coordinates}
\label{secspsy}
\label{secglobalprop}

The study of the global properties of solutions to the
spherically-symmetric Einstein--Vlasov system was initiated two
decades ago by Rein and Rendall~\cite{RR1}, cf.\ also~\cite{Rn2u2,
  Rl6}. They chose to work in coordinates where the metric takes the
form
\begin{displaymath}
  ds^{2}=-e^{2\mu(t,r)}dt^{2}+e^{2\lambda(t,r)}dr^{2}+
  r^{2}(d\theta^{2}+\sin^{2}{\theta}\,d\varphi^{2}),
\end{displaymath}
where $t\in\mathbb{R}$, $r\geq 0$, $\theta\in [0,\pi]$,
$\varphi\in [0,2\pi]$. These are called Schwarzschild
coordinates. Asymptotic flatness is expressed by the boundary
conditions
\begin{displaymath}
  \lim_{r\rightarrow\infty}\lambda(t,r)=
  \lim_{r\rightarrow\infty}\mu(t,r)=0,
  \qquad
  \forall t\geq 0.
\end{displaymath}
A regular center is also required and is guaranteed by the boundary
condition
\begin{displaymath}
  \lambda(t,0)=0
  \qquad
  \forall t\geq 0.
\end{displaymath}
The coordinates $(r,\theta,\phi)$ give rise to difficulties at $r=0$
and it is advantageous to use Cartesian coordinates. With
\begin{displaymath}
  x=(r\sin\phi\cos\theta,r\sin\phi\sin\theta,r\cos\phi)
\end{displaymath}
as spatial coordinates and
\begin{displaymath}
  v^j=p^j+(e^\lambda-1)\frac{x\cdot p}{r}\frac{x^j}{r}
\end{displaymath}
as momentum coordinates, the Einstein--Vlasov system reads
\begin{eqnarray}
&\displaystyle   \partial_{t}f+e^{\mu-\lambda}\frac{v}{\sqrt{1+|v|^{2}}}\cdot\nabla_{x}f-
  \left(\lambda_{t}\frac{x\cdot v}{r}+e^{\mu-\lambda}\mu_{r}\sqrt{1+|v|^{2}}\right)
  \frac{x}{r}\cdot\nabla_{v}f=0,\;\;\; &
  \label{Vlas}
  \\
&\displaystyle e^{-2\lambda}(2r\lambda_{r}-1)+1=8\pi r^2\rho,&
  \label{ee1}
  \\
& \displaystyle  e^{-2\lambda}(2r\mu_{r}+1)-1=8\pi r^2 p,&
  \label{ee2}
  \\
&\displaystyle \lambda_t=-4\pi\,re^{\lambda+\mu}j,&\label{ee3}
  \\
&\displaystyle e^{-2\lambda}\big(\mu_{rr}+(\mu_r-\lambda_r)(\mu_r+\frac{1}{r})\big)
-e^{-2\mu}\big(\lambda_{tt}+\lambda_t(\lambda_t-\mu_t)\big)=8\pi p_T. \label{ee4}
\end{eqnarray}
The matter quantities are defined by
\begin{eqnarray}
  \rho(t,x)&=&
  \int_{\mathbb{R}^{3}}\sqrt{1+|v|^2}f(t,x,v)\,dv,
  \label{rho}
  \\
  p(t,x)&=&\int_{\mathbb{R}^{3}}\left(\frac{x\cdot v}{r}\right)^{2}
  f(t,x,v)\frac{dv}{\sqrt{1+|v|^2}},
  \label{p}
  \\
  j(t,x)&=&\int_{\mathbb{R}^{3}}\frac{x\cdot v}{r}f(t,x,v)\,dv,
  \label{j}
  \\
  p_T(t,x)&=&\frac12\int_{\mathbb{R}^{3}}\big|\frac{x\times v}{r}\big|^2f(t,x,v)\,dv.
  \label{pT}
\end{eqnarray}
Here $\rho$ is the energy density, $j$ the current, $p$ the radial pressure, and 
$p_T$ the tangential pressure. 
Let us point out that these equations are not independent, e.g., Equations~(\ref{ee3})
and (\ref{ee4}) follow from (\ref{Vlas})--(\ref{ee2}). 

As initial data we take a spherically-symmetric, non-negative,
and continuously differentiable function $f_0$ with compact support
that satisfies
\begin{equation}
  \int_{|y|<r}\int_{\mathbb{R}^3}
  \!\!\!\sqrt{1+|v|^2}f_{0}(y,v)\,dv\,dy\,<\frac{r}{2}.
  \label{ts}
\end{equation}
This condition guarantees that no trapped surfaces are present
initially.

The set up described above is one of several possibilities. The Schwarzschild coordinates
have the advantage that the resulting system of equations can be written in a quite
condensed form. Moreover, for most initial data, 
solutions are expected to exist globally
in Schwarzschild time, which sometimes is called the polar time gauge. Let us point out here that
there are initial data leading to spacetime singularities, cf.~\cite{Rl7,AKR2u2,AR4u2}.
Hence, the question of global existence for general initial data is only relevant if
the time slicing of the spacetime is expected to be singularity avoiding, which is
the case for Schwarzschild time. We refer to~\cite{MEu2} for a general discussion on this issue.
This makes Schwarzschild coordinates tractable in the study of
the Cauchy problem. However, one disadvantage is that these coordinates only cover
a relatively small part of the spacetime, in particular trapped surfaces are not admitted. Hence,
to analyze the black-hole region of a solution these coordinates are not appropriate.
Here we only mention the other coordinates and time gauges that have been considered
in the study of the spherically symmetric Einstein--Vlasov system.
These works will be discussed in more detail in various sections below. Rendall
uses maximal-isotropic coordinates in~\cite{Rl6}. These coordinates are also considered
in~\cite{A6u2}. The Einstein--Vlasov system is investigated in double null coordinates
in~\cite{DR1, D1u2}. Maximal-areal coordinates and Eddington--Finkelstein coordinates
are used in~\cite{AR1u2, AKR1u2}, and in~\cite{AR4u2} respectively.

\subsection{Local existence and the continuation criterion}

In~\cite{RR1} it is shown that for initial data satisfying (\ref{ts})
there exists a unique, continuously-differentiable solution $f$ with
$f(0)=f_0$ on some right maximal interval $[0,T)$. If the solution
blows up in finite time, i.e., if $T<\infty$, then $\rho(t)$ becomes
unbounded as $t\rightarrow T$. Moreover, a continuation criterion is
shown that says that a local solution can be extended to a global one, 
provided $Q(t)$ can be bounded on $[0,T)$, where
\begin{equation}
\label{Qcompact}
Q(t):=\sup \{|v|:\exists (s,x)\in [0,t]\times \mathbb{R}^3 \text{ such
  that } f(s,x,v)\not= 0\}.
\end{equation}
This is
analogous to the situation for the Vlasov--Maxwell system. A control of
the $v$-support immediately implies that $\rho$ and $p$ are
bounded in view of Equations~(\ref{rho}, \ref{p}). In
the Vlasov--Maxwell case the field equations have a regularizing
effect in the sense that derivatives can be expressed through spatial
integrals, and it follows~\cite{GSt1} that the derivatives of $f$ can also
be bounded if the $v$-support is bounded. For the Einstein--Vlasov
system such a regularization is less clear, since, e.g., $\mu_r$
depends on $p$ in a point-wise manner. However, in view of Equation~(\ref{ee4})
certain combinations
of second and first order derivatives of the metric components can be
expressed in terms of the matter component $p_T$, which is a
consequence of the geodesic deviation equation. This fact turns out
to also be sufficient for obtaining bounds on the derivatives of $f$,
cf.~\cite{RR1, Rn2u2, Rl6} for details.

The local existence result discussed above holds for compactly-supported initial data.
The compact support condition in the momentum variables is in~\cite{A6u2} replaced by
the fall-off condition
\begin{equation}
\label{iddecay}
\sup_{(x,v)\in\mathbb{R}^6} (1+|v|)^5 |\mathring{f} (x,v)| < \infty.
\end{equation}
We also refer to~\cite{AR3u2} where a subclass of non-compactly-supported data is treated. 

Local existence of solutions in double null coordinates and in Eddington--Finkelstein
coordinates is established in~\cite{DR1}, and~\cite{AR4u2} respectively.

\subsection{Global existence for small initial data}
\label{globsmall}

In~\cite{RR1} the authors also consider the problem of global existence
in Schwarzschild coordinates for
small initial data for massive particles.
They show that for such data the $v$-support is
bounded on $[0,T)$. Hence, the continuation criterion implies that $T=\infty$.
The resulting spacetime in~\cite{RR1} is geodesically complete,
and the components of the energy-momentum tensor as well
as the metric quantities decay with certain algebraic
rates in $t$. The mathematical method used by Rein
and Rendall is inspired by the analogous small data result for the
Vlasov--Poisson equation by Bardos and Degond~\cite{BD}. This should
not be too surprising since for small data the gravitational fields
are expected to be small and a Newtonian spacetime should be a fair
approximation. In this context we point out that in~\cite{RR2} it is
proven that the Vlasov--Poisson system is indeed the non-relativistic
limit of the spherically-symmetric Einstein--Vlasov system, i.e., the
limit when the speed of light $c\rightarrow \infty$. In~\cite{Rl3}
this result is shown without symmetry assumptions.

As mentioned above the local and global existence problem has been studied using
other time gauges, in particular Rendall has shown global existence
for small initial data in maximal-isotropic coordinates in~\cite{Rl6}. 

The previous results refer to massive particles but they do not immediately carry over
to massless particles. This case is treated by Dafermos in~\cite{D1u2} where
global existence for small initial data is shown in double null coordinates.
The spacetimes obtained in the studies~\cite{RR1,Rl6,D1u2} are all causally
geodesically complete and appropriate decay rates of the metric and the matter  
quantities are given.

\subsection{Global existence for special classes of large initial data}
\label{globspec}

In the case of small initial data the resulting spacetime is geodesically
complete and no singularities form. A different scenario, which leads
to a future geodesically complete spacetime, is to consider initial data where
the particles are moving rapidly outwards. If the particles move sufficiently
fast the matter disperses and the gravitational attraction is not strong enough
to reverse the velocities of the particles to create a collapsing system.
This problem is studied in~\cite{AKR1u2} using a maximal time coordinate. It is
shown that the scenario described above can be realized, and that global existence holds. 

In Section~\ref{secfbh} we discuss 
results on the formation of black holes and trapped surfaces; in particular, the results
in~\cite{AKR2u2} will be presented. A corollary of the main result in~\cite{AKR2u2} 
concerns the issue of global existence and thus we mention it here. It is
shown that a particular class of initial data, which lead to formation of black holes,
have the property that the solutions exist for all Schwarzschild time. The initial data
consist of two parts: an inner part, which is a static solution of the Einstein--Vlasov
system, and an outer part with matter moving inwards. The set-up is shown to preserve
the direction of the momenta of the outer part of the matter, and it is also shown
that in Schwarzschild time the inner part and the outer part of the matter
never interact in Schwarzschild time.

\subsection{On global existence for general initial data}
\label{sec:gendat}

As was mentioned at the end of Section~\ref{secspsy}, the issue of global existence
for general initial data is only relevant in certain time gauges since there are
initial data leading to singular spacetimes. However, it is reasonable to believe
that global existence for general data may hold in a polar time gauge or a maximal time
gauge, cf.~\cite{MEu2}, 
and it is often conjectured in the literature that these time slicings are
singularity avoiding. However, there is no proof of
this statement for any matter model and it would be very satisfying to
provide an answer to this conjecture for the Einstein--Vlasov system. A 
proof of global existence in these time coordinates would also be of great importance due
to its relation to the weak cosmic censorship conjecture, 
cf.~\cite{Cu1u2,D1,DR1u2}.

The methods of proofs in the cases described in Sections~\ref{globsmall} and \ref{globspec}, 
where global existence has been shown, are all tailored to treat special classes of
initial data and they will likely not apply in more general situations.
In this section we discuss some attempts to treat general initial
data. These results are all conditional in the sense that assumptions
are made on the solutions, and not only on the initial data.

The first study on global existence for general initial data is~\cite{RRS1},
which is carried out in Schwarzschild coordinates. The authors introduce
the following variables in the momentum space adapted to spherical symmetry,
\begin{equation}
\label{wL}
  L:=|x|^2|v|^2-(x\cdot v)^2,\;\; w=\frac{x\cdot v}{r},
\end{equation}
where $L$ is the square of the angular momentum and $w$ is the radial component
of the momenta. A consequence of spherical symmetry is that angular momentum is
conserved along the characteristics. In these variables
the Vlasov equation for $f=f(t,r,w,L)$ becomes
\begin{equation}
  \partial_{t}f+e^{\mu-\lambda}\frac{w}{E}\partial_{r}f-
  \left(\lambda_{t}w+e^{\mu-\lambda}\mu_{r}E-
  e^{\mu-\lambda}\frac{L}{r^3E}\right)\partial_{w}f=0,
  \label{Vlas2}
\end{equation}
where
\begin{displaymath}
  E=E(r,w,L)=\sqrt{1+w^{2}+L/r^{2}}.
\end{displaymath}
%
%
%
%
%
%
%

The main result in~\cite{RRS1} shows that as long as there is no
matter in the ball
$$\{x\in\mathbb{R}^3:|x|\leq\epsilon\},$$
the estimate
\begin{equation}
\label{Qee}
Q(t)\leq e^{\log{Q(0)}e^{C(\epsilon)t}},
\end{equation}
holds. Here $C(\epsilon)$ is a constant, which depends on
$\epsilon$. Thus, in view of the continuation criterion this can be
viewed as a global existence result outside the center of symmetry for
initial data with compact support. This result rules out
shell-crossing singularities, which are present when, e.g., dust is
used as a matter model. The bound of $Q$ is obtained by estimating
each term individually in the characteristic equation associated with
the Vlasov equation~(\ref{Vlas2}) for the radial momentum. This
involves a particular difficulty. The Einstein equations imply that
$$\mu_r=\frac{m}{r^2}e^{2\lambda}+4\pi rpe^{2\lambda}$$
where
\begin{equation}
\label{mass}
m(t,r)=4\pi \int_0^r \eta^2\rho(t,\eta)\,d\eta,
\end{equation}
is the quasi local mass. Thus, using~(\ref{ee3}) the characteristic
equation consists of the two terms $T_1= 4\pi
re^{\mu+\lambda}(jw+pE)$, and $T_2=e^{\mu+\lambda}\frac{m}{r^2}$,
together with a term, which is independent of the matter
quantities. There is a distinct difference between the terms $T_1$ and
$T_2$ due to the fact that $m$ can be regarded as an average, since it
is given as a space integral of the energy density $\rho$, whereas $j$
and $p$ are point-wise terms. The method in~\cite{RRS1} makes use of a
cancellation property of the radial momenta in $T_1$ so that outside
the center this term is manageable but in general it seems very
unpleasant to have to treat point-wise terms of this kind. 

In~\cite{Rl6} Rendall shows global existence outside the center in
maximal-isotropic coordinates. The bound on $Q(t)$ is again obtained
by estimating each term in the characteristic equation. In this case
there are no point-wise terms in contrast to the case with
Schwarzschild coordinates. However, the terms are, in analogy with the
Schwarzschild case, strongly singular at the center.

A recent work~\cite{A6u2} gives an alternative and simplified proof of
the result in~\cite{RRS1}. In particular, the method avoids the
point-wise terms by using the fact that the characteristic system can
be written in a form such that Green's formula in the plane can be
applied. This results in a combination of terms involving second-order
derivatives, which can be substituted for by one of the Einstein
equations. This method was first introduced in~\cite{A1u2} but the
set-up is different in~\cite{A6u2} and the application of Green's
formula becomes very natural. In addition, the bound of $Q$ is
improved compared to (\ref{Qee}) and reads 
$$
Q(t)\leq Q(0)e^{C(1+t)/\epsilon t}.
$$ 
This bound is sufficient to conclude that global existence outside the
center also holds for non-compact initial data. In addition to the
global existence result outside the centre, it is shown in~\cite{A6u2}
that as long as $3m(t,r)\leq r$ and $j\leq 0$, singularities cannot
form. Note that in Schwarzschild coordinates $2m(t,r)\leq r$ always,
and that there are closed null geodesics if $3m=r$ in the
Schwarzschild static spacetime.

The method in~\cite{A6u2} also applies to the case of
maximal-isotropic coordinates studied in~\cite{Rl6}. There is an
improvement concerning the regularity of the terms that need to be
estimated to obtain global existence in the general case. A
consequence of~\cite{A6u2} is accordingly that the quite different
proofs in~\cite{RRS1} and in~\cite{Rl6} are put on the same
footing. We point out that the method can also be applied to the case
of maximal-areal coordinates. 

The results discussed above concern time gauges, which 
are expected to be singularity avoiding 
so that the issue of global existence makes sense. An 
interpretation of these results is that 
``first singularities'' (where the notion of ``first'' is tied to the causal structure), 
in the non-trapped region, must emanate from the center 
and that this case has also been shown in double null-coordinates by Dafermos 
and Rendall in~\cite{DR1}. 
The main motivation for studying
the system in these coordinates has its origin from the method of
proof of the cosmic-censorship conjecture for the Einstein--scalar
field system by Christodoulou~\cite{Cu2}. An essential part of his
method is based on the understanding of the formation of trapped
surfaces~\cite{Cu5}. 
In~\cite{D1} it is shown that a single trapped surface or marginally-trapped surface
in the maximal development implies that weak cosmic censorship holds 
The theorem holds true for any spherically-symmetric matter spacetime if
the matter model is such that ``first'' singularities necessarily
emanate from the center. The results in~\cite{RRS1} and in~\cite{Rl6} are
not sufficient for concluding that the hypothesis of the matter needed
in the theorem in~\cite{D1} is satisfied, since they concern a portion
of the maximal development covered by particular
coordinates. Therefore, Dafermos and Rendall~\cite{DR1} choose 
double-null coordinates, which cover the maximal development, and they
show that the mentioned hypothesis is satisfied for Vlasov matter.

\subsection{Self-similar solutions}

The main reason that the question of global existence in certain time
coordinates discussed in the previous Section~\ref{sec:gendat} is of
great importance is its relation to the cosmic censorship conjectures.
Now there is, in fact, no theorem in the literature, which guarantees that weak
cosmic censorship follows from such a global existence result, but there are 
strong reasons to believe that this is the case, cf.~\cite{MEu2} and \cite{AKR4u2}. 
Hence, if initial data can be constructed, which lead to naked singularities, 
then either the conjecture that global existence holds generally is false or
the viewpoint that global existence implies the absence of naked singularities
is wrong. In view of a recent result by Rendall and Velazquez~\cite{RVu2}
on self similar dust-like solutions for the massless Einstein--Vlasov system, this
issue has much current interest. Let us mention here that there is a
previous study on self-similar solutions in the massless case by
Mart{\'{\i}}n-Garc{\'{\i}}a and Gundlach~\cite{MG}. However, this
result is based on a scaling of the density function itself and
therefore makes the result less related to the Cauchy problem. Also,
their proof is, in part, based on numerics, which makes it harder to
judge the relevance of the result.

The main aim of the work~\cite{RVu2} is to establish self-similar solutions
of the massive Einstein--Vlasov system and the present result can be viewed as a
first step to achieving this. In the set-up, two simplifications are made. First,
the authors study the massless case in order
to find a scaling group, which leaves the system invariant. More precisely, the massless
system is invariant under the scaling
\[
r\to\theta r,\;\;t\to\theta t,\;\;w\to\frac{1}{\sqrt{\theta}}w,\;\;L\to\theta L.
\]
The massless assumption seems not very restrictive since, if a singularity forms,
the momenta will be large and therefore the influence of
the rest mass of the particles will be negligible, so that asymptotically
the solution can be self-similar also in the massive case, cf.~\cite{LMRu2},
for the relativistic Vlasov--Poisson system. 
The second simplification is that the possible radial momenta are restricted to
two values, which means that the density function is a distribution in this variable.
Thus, the solutions can be thought of as intermediate between
smooth solutions of the Einstein--Vlasov system and dust.

For this simplified system it turns out that the existence question of
self-similar solutions can be reduced to that of the existence of a
certain type of solution of a four-dimensional system of ordinary
differential equations depending on two parameters. The proof is based
on a shooting argument and involves relating the dynamics of solutions
of the four-dimensional system to that of solutions of certain two-
and three-dimensional systems obtained from it by limiting
processes. The reason that an ODE system is obtained is due to the
assumption on the radial momenta, and if regular initial data is
considered, an ODE system is not sufficient and a system of partial
differential equations results.

The self-similar solution obtained by Rendall and Velazquez has some
interesting properties. The solution is not asymptotically flat but there are ideas
outlined in~\cite{RVu2} of how this can be overcome. It should be pointed out here that a 
similar problem occurs in the work by Christodoulou~\cite{Cu1} for a scalar field, 
where the naked singularity solutions 
are obtained by truncating self-similar data. The singularity of the self-similar
solution by Rendall and Velazquez 
is real in the sense that the Kretschmann scalar curvature blows up.
The asymptotic structure of the solution is striking in view of
the conditional global existence result in~\cite{A6u2}. Indeed, the self similar
solution is such that $j\leq 0$, and $3m(t,r)\to r$ asymptotically, but for any $T$,
$3m(t,r)>r$ for some $t>T$. In~\cite{A6u2} global existence follows if $j\leq 0$ and if
$3m(t,r)\leq r$ for all $t$. It is also the case that if $m/r$ is close to $1/2$, then 
global existence holds in certain situations, cf.~\cite{AKR2u2}. Hence, the asymptotic
structure of the self-similar solution has properties, which have been shown to be difficult 
to treat in the search for a proof of global existence.

\subsection{Formation of black holes and trapped surfaces}
\label{secfbh}

We have previously mentioned that there exist initial data for the
spherically-symmetric Einstein--Vlasov system, which lead to formation
of black holes.

The first result in this direction was obtained by Rendall~\cite{Rl7}. 
He shows that there exist
initial data for the spherically-symmetric Einstein--Vlasov system
such that a trapped surface forms in the evolution. 
The occurrence of a trapped surface signals the formation of an event horizon. 
As mentioned above, Dafermos~\cite{D1} has proven that, if a spherically-symmetric spacetime
contains a trapped surface and the matter model satisfies certain hypotheses, 
then weak cosmic censorship holds true. In~\cite{DR1} it was then
shown that Vlasov matter does satisfy the required hypotheses. Hence, by combining
these results it follows that initial data exist, which lead to gravitational collapse
and for which weak cosmic censorship holds. However, the proof in~\cite{Rl7} 
rests on a continuity argument, and it is not possible to tell whether or not a
given initial data set will give rise to a black hole. Moreover, the mechanism 
of how trapped surfaces form is not revealed in~\cite{Rl7}.
This is in contrast  to the result in~\cite{AR4u2}, where 
explicit conditions on the initial data
are given, which guarantee the formation of trapped surfaces in the evolution.
The analysis is carried out in Eddington--Finkelstein coordinates and a central result
in~\cite{AR4u2} is to control the life span of the solution to ensure that there
is sufficient time to form a trapped surface before the solution may break down. In 
particular, weak cosmic censorship holds for these initial data.
In~\cite{AKR2u2} the formation of the event
horizon in gravitational collapse is analyzed in Schwarzschild coordinates. Note that
these coordinates do not admit trapped surfaces. The initial data in~\cite{AKR2u2}
consist of two separate parts of matter. One inner part and one outer part, in which
all particles move inward initially.
The reason for the inner part is that it is possible to choose the parameters for
the data such that the particles of the outer matter part continue to move inward
for all Schwarzschild time as long as the particles do not interact with the inner part.
This fact simplifies the analysis since the dynamics is much restricted when
the particles keep the direction of their radial momenta. The main result
is that explicit conditions on the initial data with ADM mass $M$ are given 
such that there is a family of outgoing null geodesics for which the area radius $r$ along
each geodesic is bounded by $2M$. It is furthermore
shown that if
\[
t\geq 0, \text{ and } r\geq 2M+\alpha e^{-\beta t},
\]
where $\alpha$ and $\beta$ are positive constants, then $f(t,r,\cdot,\cdot)=0$,
and the metric equals the Schwarzschild metric
\begin{equation}
\label{Schwarz}
ds^2=-\big(1-\frac{2M}{r}\big)dt^2+\big(1-\frac{2M}{r}\big)^{-1}dr^2
+r^2(d\theta^2+\sin^2{\theta}\,d\phi^2),
\end{equation}
representing a black hole with mass $M$. Hence, spacetime converges asymptotically to
the Schwarz\-schild metric.

The latter result does not reveal whether or not all the matter crosses $r=2M$ or simply
piles up at the event horizon. In~\cite{AR3u2} it is shown that for initial data, which are
closely related to those in~\cite{AKR2u2}, but such that the radial momenta are
unbounded, all the matter do cross the event horizon asymptotically in Schwarzschild time.
This is in contrast to what happens to freely-falling observers in a static
Schwarzschild spacetime, since they will never reach the event horizon. 

The result in~\cite{AKR2u2} is reconsidered in~\cite{AKR3u2}, where an additional argument is
given to match the definition of weak cosmic censorship given in~\cite{Cu1u2}. 

It is natural to relate the results of~\cite{AKR2u2,AR4u2} to those of Christodoulou
on the spherically-symmetric Einstein-scalar field system~\cite{Cu2u2} and \cite{Cu5}.
In~\cite{Cu2u2} it is shown that if the final Bondi mass $M$ is different from zero,
the region exterior to the sphere $r=2M$ tends to the Schwarzschild metric
with mass $M$ similar to the result in~\cite{AKR2u2}.
In~\cite{Cu5} explicit conditions on the initial data are specified,
which guarantee the formation of trapped surfaces. This paper played a crucial role in
Christodoulou's proof~\cite{Cu2} of the weak and strong cosmic censorship conjectures.
The conditions on the initial data in~\cite{Cu5} allow the ratio
of the Hawking mass and the area radius to cover the full range, i.e., $2m/r\in (0,1)$,
whereas the conditions in~\cite{AR4u2} require $2m/r$ to be close to one. Hence, 
it would be desirable to improve the conditions on the initial data in~\cite{AR4u2}, 
although the conditions by Christodoulou for a scalar field are not expected to be 
sufficient in the case of Vlasov matter. 

\subsection{Numerical studies on critical collapse}

In~\cite{RRS2} a numerical study on critical collapse for the Einstein--Vlasov system
was initiated. A numerical scheme originally used for the Vlasov--Poisson system was
modified to the spherically-symmetric Einstein--Vlasov system. It has
been shown by Rein and Rodewis~\cite{RRo} that the numerical scheme
has desirable convergence properties. (In the Vlasov--Poisson case,
convergence was proven in~\cite{Sc3}, see also~\cite{GV}).

The speculation discussed above that there may be no naked singularities formed for
any regular initial data is in part based on
the fact that the naked singularities that occur in
scalar field collapse appear to be associated with the existence of
type~II critical collapse, while Vlasov matter is of type I.
The primary goal in~\cite{RRS2} was indeed to decide whether Vlasov
matter is type~I or type II.


These different types of matter are defined as follows. Given small
initial data, no black holes form and matter will
disperse. For large data, black holes will form and
consequently there is a transition regime separating dispersion of
matter and formation of black holes. If we introduce a parameter $A$
on the initial data such that for small $A$ dispersion occurs and for
large $A$ a black hole is formed, we get a critical value $A_\mathrm{c}$
separating these regions. If we take $A>A_\mathrm{c}$ and denote by
$m_\mathrm{B}(A)$ the mass of the black hole, then if
$m_\mathrm{B}(A)\rightarrow 0$ as $A\rightarrow A_\mathrm{c}$ we have type
II matter, whereas for type I matter this limit is positive and there
is a mass gap. For more information on critical collapse we refer to
the review paper by Gundlach~\cite{Gu}.

The conclusion of~\cite{RRS2} is that Vlasov matter is of type I.
There are two other independent numerical simulations on critical collapse
for Vlasov matter~\cite{OC,AR1u2}. In these simulations, maximal-areal coordinates
are used rather than Schwarzschild coordinates as in~\cite{RRS2}.
The conclusion of these studies agrees with the one in~\cite{RRS2}.


\subsection{The charged case}

We end this section with a discussion of the spherically-symmetric
Einstein--Vlasov--Maxwell system, i.e., the case considered above with
charged particles. Whereas the constraint equations in the uncharged case,
written in Schwarzschild coordinates, do not involve solving any difficulties
once the distribution function is given, the charged case is more challenging.
However, in~\cite{NNR} it is shown that solutions to the constraint
equations do exist for the Einstein--Vlasov--Maxwell system.
In~\cite{NN} local existence is shown together with a
continuation criterion, and global existence for small initial data
is shown in~\cite{Nuu2}.

\newpage

\section{The Cosmological Cauchy Problem}
\label{sec:cosmological-cauchy-problem}

In this section we discuss the Einstein--Vlasov system for cosmological 
spacetimes, i.e., spacetimes that possess a compact Cauchy surface.
The ``particles'' in the kinetic description are in this case galaxies 
or even clusters of galaxies. The main goal is to determine 
the global properties of the solutions
to the Einstein--Vlasov system for initial data given on a compact 3-manifold. 
In order to do so, a global time
coordinate $t$ must be found and the asymptotic
behavior of the solutions when $t$ tends to its limiting values has
to be analyzed. This might correspond to approaching a singularity, 
e.g., the big bang singularity, or to a phase of unending expansion.

Presently, the aim of most of the studies of the cosmological Cauchy problem
has been to show existence for unrestricted initial data and the results that
have been obtained are in cases with symmetry (see, however,~\cite{Ae}, where
to some extent global properties are shown in the case without symmetry).
These studies will be reviewed below.
A recent and very extensive work by Ringstr\"{o}m
has, on the other hand, a different aim, i.e., to show stability of homogeneous
cosmological models, and concerns the general case without symmetry.
The size of the Cauchy data is in this case very restricted 
but, since Ringstr\"{o}m allows general perturbations, there are no symmetries available 
to reduce the complexity of the Einstein--Vlasov system. This result will be reviewed 
at the end of this section.

\subsection{Spatially-homogeneous spacetimes}

The only spatially-homogeneous spacetimes admitting a compact Cauchy
surface are the Bianchi types I, IX and the Kantowski-Sachs model; to
allow for cosmological solutions with more general symmetry types, it is
enough to replace the condition that the spacetime is spatially
homogeneous, with the condition that the universal covering of spacetime
is spatially homogeneous. Spacetimes with this property are called locally
spatially homogeneous and these include, in addition, the Bianchi
types II, III, V, VI\sub{0}, VII\sub{0}, and VIII.

One of the first studies on the Einstein--Vlasov system for
spatially-homogeneous spacetimes is the work~\cite{Rl4} by Rendall. He
chooses a Gaussian time coordinate and investigates the maximal range
of this time coordinate for solutions evolving from homogeneous data.
For Bianchi~IX and for Kantowski--Sachs spacetimes he finds that the range is finite
and that there is a curvature singularity in both the past and the future time directions.
For the other Bianchi types there is a curvature singularity in the past,
and to the future spacetime is causally geodesically complete. In particular,
strong cosmic censorship holds in these cases.

Although the questions on curvature singularities and geodesic completeness are very
important, it is also desirable to have more detailed information on the asymptotic
behavior of the solutions, and, in particular, to understand in which situations
the choice of matter model is essential for the asymptotics.

In recent years several studies on the Einstein--Vlasov system for spatially locally
homogeneous spacetimes have been carried out with the goal to obtain
a deeper understanding of the asymptotic structure of the solutions. Roughly,
these investigations can be divided into two cases: (i) studies on non-locally
rotationally symmetric (non-LRS) Bianchi~I models and (ii) studies of LRS
Bianchi models.

In case (i) Rendall shows in~\cite{Rl1u2} that solutions converge to dust
solutions for late times. Under the additional assumption of small initial data
this result is extended by Nungesser~\cite{Nru2}, who gives the rate
of convergence of the involved quantities.
In~\cite{Rl1u2} Rendall also raises the question of the existance of solutions with complicated
oscillatory behavior
towards the initial singularity may exist for Vlasov matter,
in contrast to perfect fluid matter. Note that for a perfect fluid
the pressure is isotropic, whereas for Vlasov matter the pressure may be anisotropic, and this
fact could be sufficient to drastically change the dynamics.
This question is answered in~\cite{HUu2}, where the existence of a heteroclinic
network is established as a possible asymptotic state. This implies a complicated oscillating
behavior, which differs from the dynamics of perfect fluid solutions. The results
in~\cite{HUu2} were then put in a more general context by Calogero and Heinzle~\cite{CH1u2},
where quite general anisotropic matter models are considered.

In case (ii) the asymptotic
behaviour of solutions has been
analyzed in~\cite{RT,RU,CH2u2,CH3u2}. In~\cite{RT}, the case of
massless particles is considered, whereas the massive case is studied
in~\cite{RU}. Both the nature of the initial singularity and the phase
of unlimited expansion are analyzed. The main concern in these two works is the behavior
of Bianchi models I, II, and III. The authors compare their solutions
with the solutions to the corresponding perfect fluid models. A
general conclusion is that the choice of matter model is very
important since, for all symmetry classes studied, there are differences
between the collision-less model and a perfect fluid model, both
regarding the initial singularity and the expanding phase. The most
striking example is for the Bianchi~II models, where they find
persistent oscillatory behavior near the singularity, which is quite
different from the known behavior of Bianchi type II perfect fluid
models. In~\cite{RU} it is also shown that solutions for massive
particles are asymptotic to solutions with massless particles near the
initial singularity. For Bianchi~I and II, it is also proven that
solutions with massive particles are asymptotic to dust solutions at
late times. It is conjectured that the same also holds true for
Bianchi~III. This problem is then settled by Rendall in~\cite{Rl8}.
The investigation~\cite{CH2u2} concerns a large class of anisotropic matter
models, and, in particular, it is shown that solutions of the Einstein--Vlasov system
with massless particles oscillate in the limit towards the past singularity for
Bianchi~IX models. This result is extended to the massive case in~\cite{CH3u2}.

Before finishing this section we mention two other investigations on homogeneous models
with Vlasov matter. In~\cite{Le1} Lee considers the homogeneous spacetimes with
a cosmological constant for all Bianchi models
except Bianchi type IX. She shows
global existence as well as future causal geodesic completeness. She also obtains the time
decay of the components of the energy momentum tensor as
$t\to\infty$, and she shows that spacetime is asymptotically dust-like. Anguige~\cite{Aeu2}
studies the conformal Einstein--Vlasov system for massless particles, which admit
an isotropic singularity. He shows that the Cauchy problem is well posed with
data consisting of the limiting density function at the singularity.

\subsection{Inhomogeneous models with symmetry}

In the spatially homogeneous case the metric can be written in a form that
is independent of the spatial variables and this leads to an enormous
simplification. Another class of spacetimes that are highly symmetric but require
the metric to be spatially dependent are those that admit a group of isometries
acting on two-dimensional spacelike orbits, at least after
passing to a covering manifold. The group may be two-dimensional
(local $U(1)\times U(1)$ or \T{2} symmetry) or three-dimensional
(spherical, plane, or hyperbolic symmetry).
In all these cases, the quotient of spacetime by the
symmetry group has the structure of a two-dimensional Lorentzian
manifold $Q$. The orbits of the group action (or appropriate quotients
in the case of a local symmetry) are called surfaces of
symmetry. Thus, there is a one-to-one correspondence between surfaces
of symmetry and points of $Q$. There is a major difference between the
cases where the symmetry group is two- or three-dimensional. In the
three-dimensional case no gravitational waves are admitted, in
contrast to the two-dimensional case where the evolution part
of the Einstein equations are non-linear wave equations.

Three types of time coordinates
that have been studied in the inhomogeneous case are CMC, areal, and
conformal coordinates. A CMC time coordinate $t$ is one where each
hypersurface of constant time has constant mean curvature and on
each hypersurface of this kind the value of $t$ is the mean curvature
of that slice. In the case of areal coordinates, the time coordinate
is a function of the area of the surfaces of symmetry, e.g.,
proportional to the area or proportional to the square root of
the area. In the case of conformal coordinates, the metric on the
quotient manifold $Q$ is conformally flat. The CMC
and the areal coordinate foliations are both geometrically-based time
foliations. The advantage with a CMC approach is that the definition
of a CMC hypersurface does not depend on any symmetry assumptions and
it is possible that CMC foliations will exist for general
spacetimes. The areal coordinate foliation, on the other hand, is
adapted to the symmetry of spacetime but it has analytical advantages
and detailed information about the asymptotics can be derived. 
The conformal coordinates have mainly served as a useful framework for 
the analysis to obtain geometrically-based time foliations. 

\subsubsection{Surface symmetric spacetimes}
\label{sec:surface-symmetric-spacetimes}

Let us now consider spacetimes $(M,g)$ admitting a three-dimensional
group of isometries. The topology of $M$ is assumed to be
$\mathbb{R}\times S^1\times F$, with $F$ a compact two-dimensional
manifold. The universal covering $\hat{F}$ of $F$ induces a spacetime
$(\hat{M},\hat{g})$ by $\hat{M}=\mathbb{R}\times S^1\times\hat{F}$ and
$\hat{g}=p^{*}g$, where $p:\hat{M}\rightarrow M$ is the canonical
projection. A three-dimensional group $G$ of isometries is assumed to
act on $(\hat{M},\hat{g})$. If $F=S^2$ and $G=SO(3)$, then $(M,g)$ is
called spherically symmetric; if $F=T^2$ and $G=E_2$ (Euclidean
group), then $(M,g)$ is called plane symmetric; and if $F$ has genus
greater than one and the connected component of the symmetry group $G$
of the hyperbolic plane $H^2$ acts isometrically on $\hat{F}=H^2$,
then $(M,g)$ is said to have hyperbolic symmetry.

In the case of spherical symmetry the existence of one compact CMC
hypersurface implies that the whole spacetime can be covered by a CMC
time coordinate that takes all real values~\cite{Rl1,BR}. The
existence of one compact CMC hypersurface in this case was proven
by Henkel~\cite{He1} using the concept of prescribed mean
curvature (PMC) foliation. Accordingly, this gives a complete picture
in the spherically symmetric case regarding CMC foliations.
In the case of areal coordinates, Rein~\cite{Rn2} has shown, under a size
restriction on the initial data, that the past of an initial
hypersurface can be covered, and that the Kretschmann scalar blows up.
Hence, the initial singularity for the restricted data is both a
crushing and a curvature singularity.
In the future direction it is shown that
areal coordinates break down in finite time.

In the case of plane and hyperbolic symmetry, global existence to the past
was shown by Rendall~\cite{Rl1} in CMC time. This implies that the past singularity is a
crushing singularity since the mean curvature blows up at the singularity.
Also in these cases Rein showed~\cite{Rn2} under a size restriction on the initial data,
that global existence to the past in areal time and blow up of the Kretschmann scalar
curvature as the singularity is approached.
Hence, the singularity is both a crushing and a curvature singularity in these cases too.
In both of these works the question of global existence
to the future was left open.
This gap was closed in~\cite{ARR}, and global existence to the future was
established in both CMC and areal time coordinates.
The global existence result for CMC time is a consequence of the global
existence theorem in areal coordinates, together with a theorem by
Henkel~\cite{He1} which shows that there exists at least one
hypersurface with (negative) constant mean curvature. In addition, the past
direction is analyzed in~\cite{ARR} using areal coordinates, and global existence is 
shown without a size restriction on the data.
It is not concluded if the past singularity, without the smallness condition on the data,
is a curvature singularity as well. The issues discussed above have also been studied
in the presence of a cosmological constant, cf.~\cite{T,TN}. In particular it is shown
that in the spherically-symmetric case, if $\Lambda>0$,
global existence to the future holds in areal time for some special classes of initial data,
which is in contrast to the case with $\Lambda=0$. In this context we also mention
that surface symmetric spacetimes with Vlasov matter and with a Maxwell field have
been investigated in~\cite{Tcu2}.

An interesting question, which essentially was left open in the studies mentioned
above, is whether the areal time coordinate,
which is positive by definition, takes all values in the range
$(0,\infty)$ or only in $(R_0,\infty)$ for some positive $R_0$.
It should here be pointed out that there is an example for vacuum spacetimes with \T{2}
symmetry (which includes the plane symmetric case) where $R_0>0$. This
question was first resolved by Weaver~\cite{Wea} for \T{2} symmetric spacetimes
with Vlasov matter. Her result
shows that if spacetime contains Vlasov matter ($f\not= 0$) then
$R_0=0$. Smulevici~\cite{S2u2} has recently shown, under the same
assumption, that $R_0=0$ also in the hyperbolic case. Smulevici also
includes a cosmological constant $\Lambda$ and shows that both the
results, for plane (or \T{2}) symmetry and hyperbolic symmetry, are
valid for $\Lambda\geq 0$.

The important question of strong cosmic censorship for surface-symmetric spacetimes has
recently been investigated by neat methods 
by Dafermos and Rendall~\cite{DR3u2,DR1u2}. The standard
strategy to show cosmic censorship is to either show causal geodesic completeness in case
there are no singularities, or to show that some curvature invariant blows up
along any incomplete causal geodesic. In both cases no causal geodesic can leave
the maximal Cauchy development in any extension if we assume that the extension
is $C^2$. In~\cite{DR3u2,DR1u2} two alternative approaches are investigated.
Both of the methods rely on the symmetries of the spacetime. The first method is independent
of the matter model and exploits a rigidity property of Cauchy horizons inherited
from the Killing fields. The areal time described above is defined in terms
of the Killing fields and a consequence of the method by Dafermos and Rendall is
that the Killing fields extend continuously to a Cauchy horizon, if one exists.
Now, since global existence has been shown in areal time it follows that there cannot
be an extension of the maximal hyperbolic development to the future. This
method is useful for the expanding future direction. The second method is dependent
on Vlasov matter and the idea is to follow the trajectory of a particle, which crosses
the Cauchy horizon and shows that the conservation laws for the particle motion
associated with the symmetries of the spacetime,
such as the angular momentum, lead to a contradiction. In most of the cases
considered in~\cite{DR3u2} there is an assumption on the initial data for 
the Vlasov equation, which implies
that the data have non-compact support in the momentum space. It 
would be desirable to relax this assumption. 
The results of the studies~\cite{DR3u2,DR1u2} can be summarized as follows. For plane
and hyperbolic symmetry strong cosmic censorship is shown when $\Lambda\geq 0$.
The restriction that matter has non-compact support in the momentum space is here imposed
except in the plane case with $\Lambda=0$.
In the spherically-symmetric case cosmic
censorship is shown when $\Lambda=0$. In the case of $\Lambda>0$ a detailed geometric
characterization of possible boundary components of spacetime is given. The difficulties
to show cosmic censorship in this case are related to possible formation of
extremal Schwarzschild-de-Sitter-type black holes. Cosmic censorship in the past direction
is also shown for all symmetry classes, and for all values of $\Lambda$, for a special 
class of anti-trapped initial data. 

Although the methods developed in~\cite{DR3u2,DR1u2} provide
a lot of information on the asymptotic structure of the solutions, questions on
geodesic completeness and curvature blow up are not answered. In a few
cases, information on these issues has been obtained. As mentioned above, blow up
of the Kretschmann scalar curvature has been shown for restricted initial data~\cite{Rn2}.
In the case of hyperbolic
symmetry causal future geodesic completeness has been established by Rein~\cite{Rn5}
when the initial data are small.
The plane and hyperbolic symmetric cases with a
positive cosmological constant are analyzed in~\cite{TR}.
The authors show global existence to the future in areal time, and in particular
they show that the spacetimes are future geodesically complete. The positivity
of the cosmological constant is crucial for the latter result.
A form of the cosmic no-hair conjecture is also obtained in~\cite{TR}. It is shown
that the de Sitter solution acts as a model for the dynamics of the solutions by proving
that the generalized Kasner exponents tend to $1/3$ as $t\to\infty$,
which in the plane case is the de Sitter solution.

\subsubsection{Gowdy and \T{2} symmetric spacetimes}

The first study of spacetimes admitting a two-dimensional isometry group was 
carried out by Rendall~\cite{Rl2} in the case of local
\T{2} symmetry. For a discussion
of the possible topologies of these spacetimes we refer to the original
paper. In the model case the spacetime is topologically of the form
$\mathbb{R}\times T^3$, and to simplify our discussion later on we
write down the metric in areal coordinates for this type of spacetime:
\begin{eqnarray}
  g&=&e^{2(\eta-U)}(-\alpha \, dt^{2}+d\theta^{2})
  +e^{-2U}t^{2}[dy+H\,d\theta+M\,dt]^{2}
  \nonumber
  \\
  &&+e^{2U}[dx+A\,dy+(G+AH)\,d\theta+(L+AM)\,dt]^{2}.
  \label{areal}
\end{eqnarray}
Here the metric coefficients $\eta$, $U$, $\alpha$, $A$, $H$, $L$, and
$M$ depend on $t$ and $\theta$ and $\theta,x,y\in S^1$. In~\cite{Rl2}
CMC coordinates are in fact considered rather than areal coordinates.
Under the hypothesis that there exists at least one CMC hypersurface,
Rendall proves for general initial data
that the past of the given CMC hypersurface can be globally foliated by CMC
hypersurfaces and that the mean curvature of these hypersurfaces blows
up at the past singularity. The future direction was left open.
The result in~\cite{Rl2} holds for Vlasov matter and for matter
described by a wave map. That the choice of matter model is important was shown
in~\cite{Rl5}, where a non-global existence result for dust is given,
which leads to examples of spacetimes~\cite{IsR} that are not covered
by a CMC foliation.

There are several possible subcases to the \T{2} symmetric
class. The plane case, where the symmetry group is three-dimensional, is
one subcase and the form of the metric in areal coordinates is
obtained by letting $A=G=H=L=M=0$ and $U=\log{t}/2$ in
Equation~(\ref{areal}). Another subcase, which still admits only two Killing
fields (and which includes plane symmetry as a special case), is Gowdy
symmetry. It is obtained by letting $G=H=L=M=0$ in
Equation~(\ref{areal}). In~\cite{An4} Gowdy symmetric
spacetimes with Vlasov matter are considered, and it is proven that the entire maximal
globally hyperbolic spacetime can be foliated by constant areal time
slices for general initial data. The areal coordinates are
used in a direct way for showing global existence to the future,
whereas the analysis for the past direction is carried out in
conformal coordinates. These coordinates are not fixed to the geometry
of spacetime and it is not clear that the entire past has been
covered. A chain of geometrical arguments then shows that areal
coordinates indeed cover the entire spacetime.
The method in~\cite{An4} was in turn inspired by the work~\cite{BCIM} for
vacuum spacetimes, where the idea of using conformal coordinates in the
past direction was introduced.
As pointed out in~\cite{ARR}, the
result by Henkel~\cite{He2} guarantees the existence of one CMC
hypersurface in the Gowdy case and, together with the global areal
foliation in~\cite{An4}, it follows that Gowdy spacetimes with Vlasov
matter can be globally covered by CMC hypersurfaces as well.
The more general case of \T{2} symmetry was
considered in~\cite{ARW}, where global
CMC and areal time foliations were established for general initial data.
In these results, the question whether or not the areal time coordinate takes values in
$(0,\infty)$ or in $(R_0,\infty)$, $R_0>0$, was left open. As we pointed
out in Section~\ref{sec:surface-symmetric-spacetimes}, this issue was
solved by Weaver~\cite{Wea} for \T{2} symmetric spacetimes
with the conclusion that $R_0=0$, if the density function $f$
is not identically zero initially. In the case of \T{2} symmetric
spacetimes, with a positive cosmological constant, Smulevici~\cite{S2u2} has shown
global existence in areal time with the property that $t\in (0,\infty)$.

The issue of strong cosmic censorship for \T{2} symmetric spacetimes has been
studied by Dafermos and Rendall using the methods, which were developed in
the surface symmetric case described above. 
In~\cite{DR2u2} strong cosmic censorship is shown under the same restriction on
the initial data that was imposed in the surface symmetric case, which implies
that the data have non-compact support in the momentum variable. Their result 
has been extended to the case with a positive cosmological constant by Smulevici~\cite{S1u2}.


\subsection{Cosmological models with a scalar field}
\label{subsection_2_3}

The present cosmological observations indicate that the expansion of
the universe is accelerating, and this has influenced theoretical
studies in the field during the last decade. One way to produce models
with accelerated expansion is to choose a positive cosmological
constant. Another way is to include a
non-linear scalar field among the matter fields, and
in this section we review the results for the Einstein--Vlasov system, where
a linear or non-linear scalar field have been included into the model.

Lee considers in~\cite{Le2} the case where a non-linear scalar field
is coupled to Vlasov matter. The form of the energy momentum tensor
then reads
\begin{equation}
  T_{\alpha\beta}=T_{\alpha\beta}^\mathrm{Vlasov}+
  \nabla_{\alpha}\phi\nabla_{\beta}\phi-
  \left(\frac{1}{2}\nabla^{\gamma}\phi\nabla_{\gamma}\phi+V(\phi)\right)
  g_{\alpha\beta}.
  \label{TVNS}
\end{equation}
Here $\phi$ is the scalar field and $V$ is a potential, and the
Bianchi identities lead to the following equation for the scalar
field:
\begin{equation}
  \nabla^{\gamma}\nabla_{\gamma}\phi=V'(\phi).
  \label{pot}
\end{equation}
Under the assumption that $V$ is a non-negative $C^2$ function, global
existence to the future is obtained, and if the potential is restricted
to the form
\begin{displaymath}
  V(\phi)=V_{0}e^{-c\Phi},
\end{displaymath}
where $0<c<4\sqrt{\pi}$, then future geodesic completeness is proven.

In~\cite{TNR} the Einstein--Vlasov system with a linear scalar field
is analyzed in the case of plane, spherical, and hyperbolic
symmetry. Here, the potential $V$ in Equations~(\ref{TVNS}) and (\ref{pot})
is zero. A local existence theorem and a continuation criterion,
involving bounds on derivatives of the scalar field in addition to a
bound on the support of one of the moment variables, is proven. For
the Einstein scalar field system, i.e., when $f=0$, the continuation
criterion is shown to be satisfied in the future direction, and global
existence follows in that case. The work~\cite{Tu2} extends the result 
in the plane and hyperbolic case to a global result in the future direction. 
In the plane case when $f=0,$ the solutions are shown to be future geodesically 
complete. The past time direction is considered in~\cite{TRu2} and global existence 
is proven. It is also shown that the singularity is crushing and
that the Kretschmann scalar diverges uniformly as the singularity is approached.

\subsection{Stability of some cosmological models}

In standard cosmology, the universe is taken to be spatially homogeneous and
isotropic. This is a strong assumption leading to severe restrictions of the possible
geometries as well as of the topologies of the universe. Thus, it is natural to ask
if small perturbations of an initial data set, which corresponds to an expanding model of
the standard type, give rise to solutions that are similar globally to the future?

In a recent work, Ringstr\"{o}m~\cite{Riu2} considers the Einstein--Vlasov system and he gives
an affirmative answer to the stability question for some of the standard cosmologies.


The standard model of the universe is spatially homogeneous and isotropic,
has flat spatial hypersurfaces of homogeneity, a positive cosmological
constant and the matter content consists of a radiation fluid and dust.
Hence, to investigate the question on stability it is natural to consider cosmological
solutions with perfect fluid matter and a positive cosmological constant.
However, as is shown by Ringstr\"{o}m,
the standard model can be well approximated by a solution of the Einstein--Vlasov
system with a positive cosmological constant. Approximating dust with Vlasov
matter is straightforward, whereas approximating a radiation fluid is
not. By choosing the initial support of the distribution function suitably,
Ringstr\"{o}m shows that Vlasov matter can be made to mimic a radiation fluid
for a prescribed amount of time; sooner or later the matter will behave like dust,
but the time at which the approximation breaks down can be chosen to be large enough that
the radiation is irrelevant to the future of that time in the standard picture.

The main results in~\cite{Riu2} are stability of expanding, spatially compact, spatially
locally homogeneous solutions to the Einstein--Vlasov system with a positive
cosmological constant as well as a construction of solutions with arbitrary compact
spatial topology. In other words, the assumption of almost spatial homogeneity and isotropy
does not seem to impose a restriction on the allowed spatial topologies.

Let us mention here some related works although these do not concern the Einstein--Vlasov 
system. Ringstr\"{o}m considers the case where the matter model is 
a non-linear scalar field in~\cite{Ri1u2} and \cite{Ri2u2}. The background solutions, 
which Ringstr\"{o}m perturb and which are shown to be stable, have accelerated expansion. 
In~\cite{Ri1u2} the expansion is exponential and in~\cite{Ri2u2} it is of power law type. 
The corresponding problem for a fluid has been treated in~\cite{RSu2} and \cite{Spu2}, 
and the Newtonian case is investigated in~\cite{Lehou2} and \cite{BRRu2} 
for Vlasov and fluid matter respectively.

\newpage
\section{Stationary Asymptotically-Flat Solutions}

Equilibrium states in galactic dynamics can be described as static or stationary
solutions of the Einstein--Vlasov system, or of the Vlasov--Poisson
system in the Newtonian case. Here we consider the relativistic case
and we refer to the excellent review paper~\cite{Rn1u2} for
the Newtonian case. First, we discuss spherically-symmetric solutions for which 
the structure is quite well understood. On the other hand, 
almost nothing is known about the stability of the spherically-symmetric static solutions
of the Einstein--Vlasov system, which is in
sharp contrast to the situation for the Vlasov--Poisson system. At the end
of this section a recent result~\cite{AKR4u2} on axisymmetric static solutions will
be presented.

\subsection{Existence of spherically-symmetric static solutions}
\label{existssss}

Let us assume that spacetime is static and spherically symmetric. 
Let the metric have the form
\begin{displaymath}
  ds^{2}=-e^{2\mu(r)}\,dt^{2}+e^{2\lambda(r)}\,dr^{2}
  +r^{2}(d\theta^{2}+\sin^{2}{\theta}\,d\varphi^{2}),
\end{displaymath}
where $r\geq 0$, $\theta\in [0,\pi]$, $\varphi\in [0,2\pi]$. As before,
asymptotic flatness is expressed by the boundary conditions
\begin{displaymath}
  \lim_{r\rightarrow\infty}\lambda(r)=
  \lim_{r\rightarrow\infty}\mu(r)=0,
\end{displaymath}
and a regular center requires 
\begin{displaymath}
  \lambda(0)=0.
\end{displaymath}
Following the notation in Section~\ref{secspsy}, the time-independent 
Einstein--Vlasov system reads
\begin{eqnarray}
  e^{\mu-\lambda}\frac{v}{\sqrt{1+|v|^2}}\cdot\nabla_{x}f
  -\sqrt{1+|v|^2}e^{\mu-\lambda}\mu_{r}
  \frac{x}{r}\cdot\nabla_{v}f&=&0,
  \label{Vlas3}
  \\
  e^{-2\lambda}(2r\lambda_{r}-1)+1&=&8\pi r^2\rho,
  \label{ee12}
  \\
  e^{-2\lambda}(2r\mu_{r}+1)-1&=&8\pi r^2 p.
  \label{ee22}
\end{eqnarray}
Recall that there is an additional Equation~(\ref{ee4}) of second order, 
which contains the tangential pressure $p_T$, but we leave it out since 
it follows from the equations above. 
The matter quantities are defined as before:
\begin{eqnarray}
  \rho(x)&=&
  \int_{\mathbb{R}^{3}}\sqrt{1+|v|^{2}}f(x,v)\,dv,
  \label{rho3}
  \\
  p(x)&=&\int_{\mathbb{R}^{3}}\left(\frac{x\cdot v}{r}\right)^{2}
  f(x,v)\frac{dv}{\sqrt{1+|v|^{2}}}.
  \label{p3}
\end{eqnarray}
The quantities
\begin{displaymath}
  E:=e^{\mu(r)}\sqrt{1+|v|^{2}},
  \qquad
  L=|x|^2|v|^{2}-(x\cdot v)^2=|x\times v|^2,
\end{displaymath}
are conserved along characteristics. $E$ is the particle energy and
$L$ is the angular momentum squared. If we let
\begin{equation}
\label{ansatzphi}
  f(x,v)=\Phi(E,L),
\end{equation}
for some function $\Phi$, the Vlasov equation is automatically satisfied. 

A common assumption in the literature is to restrict the form of $\Phi$ to
\begin{equation}
  \Phi(E,L)=\phi(E)(L-L_0)_{+}^l,
  \label{pol}
\end{equation}
where $l>-1/2,\,L_0\geq 0$ and $x_+=\max\{x,0\}$. If we furthermore
assume that 
\[
\phi(E)=(E-E_0)^k_{+},\;k>-1,
\]
for some positive constant $E_0$, then we obtain the {\it polytropic ansatz}.
The case of isotropic pressure is obtained by letting $l=0$ 
so that $\Phi$ only depends on $E$.

In passing, we mention that for the Vlasov--Poisson system it has been
shown~\cite{BFH} that every static spherically-symmetric solution must
have the form $f=\Phi(E,L)$. This is referred to as Jeans' theorem. It
was an open question for some time whether or not this was
also true for the Einstein--Vlasov system.  This was settled in 1999
by Schaeffer~\cite{Sc4}, who found solutions that do not have this
particular form globally on phase space, and consequently, Jeans'
theorem is not valid in the relativistic case. However, almost all
results on static solutions are based on this ansatz.

By inserting the ansatz $f(x,v)=\Phi(E,L)$ in the matter quantities $\rho$ and $p$,
a non-linear system for $\lambda$ and $\mu$ is obtained, in which
\begin{eqnarray*}
  e^{-2\lambda}(2r\lambda_{r}-1)+1&=&8\pi r^2G_{\Phi}(r,\mu),
  \\
  e^{-2\lambda}(2r\mu_{r}+1)-1&=&8\pi r^2 H_{\Phi}(r,\mu),
\end{eqnarray*}
where
\begin{eqnarray*}
  G_{\Phi}(r,\mu)&=&\frac{2\pi}{r^2}\int_1^{\infty}\!\!
  \int_0^{r^2(\epsilon^2-1)}\!\!\Phi(e^{\mu(r)}\epsilon,L)
  \frac{\epsilon^2}{\sqrt{\epsilon^2-1-L/r^2}}\,dL\,d\epsilon,
  \\
  H_{\Phi}(r,\mu)&=&\frac{2\pi}{r^2}\int_1^{\infty}\!\!
  \int_0^{r^2(\epsilon^2-1)}\!\!\Phi(e^{\mu(r)}\epsilon,L)
  \sqrt{\epsilon^2-1-L/r^2}\,dL\,d\epsilon.
\end{eqnarray*}

Existence of solutions to this system was first proven in the case of
isotropic pressure in~\cite{RR3}, and extended to anisotropic pressure in~\cite{Rn3}.
The main difficulty is to show that the solutions have finite ADM mass and compact support. 
The argument in these works to obtain a solution of compact support
is to perturb a steady state of the Vlasov--Poisson system, which is
known to have compact support. 
Two different types of solutions are constructed, those with a
regular centre~\cite{RR3, Rn3}, and those with a Schwarzschild
singularity in the centre~\cite{Rn3}. 

In~\cite{RR4}, Rein and Rendall 
go beyond the polytropic ansatz and obtain steady states with compact support 
and finite mass under the assumption that $\Phi$ satisfies 
\begin{displaymath}
  \Phi(E,L)=c(E-E_0)^k_{+}L^l+O((E_0-E)_{+}^{\delta+k})L^l
  \text{ as } E\rightarrow E_0,
\end{displaymath}
where $k>-1$, $l>-1/2$, $k+l+1/2>0$, $k<l+3/2$. 
This result is obtained in a more direct way and
is not based on the perturbation argument used in~\cite{RR3,Rn3}. Their 
method is inspired by a work on stellar models by Makino~\cite{M}, in
which he considers steady states of the Euler--Einstein
system. In~\cite{RR4} there is also a discussion about
steady states that appear in the astrophysics literature, and it is shown 
that their result applies to most of these steady states. 
An alternative method to obtain steady states with finite
radius and finite mass, which is based on a dynamical system analysis,
is given in~\cite{FHUu2}.

\subsection{The structure of spherically-symmetric steady states}

All solutions described so far have the property that the support of
$\rho$ contains a ball about the centre. In~\cite{Rn4} Rein showed that
steady states also exist whose support is a finite, spherically-symmetric shell with a vacuum region in the center. In~\cite{A3u2} it
was shown that there are shell solutions, which have an arbitrarily
thin thickness. A systematic
study of the structure of spherically-symmetric static solutions
was carried out mainly by numerical means in~\cite{AR2u2} and we
now present the conclusions of this investigation.

By prescribing the value $\mu(0)$, the equations can be solved, but the resulting solution
will in general not satisfy the boundary condition $\mu(\infty)=0$, but it will have
some finite limit. It is then
possible to shift both the cut-off energy $E_0$ and the solution by this limit to
obtain a solution, which satisfies $\mu(\infty)=0$. A convenient way to handle
the problem that $E_0$ and $\mu(0)$ cannot both be treated as free parameters
is to use the ansatz
\[
f(x,v)=(1-E/E_0)^k_+(L-L_0)^l_+,\;k\geq 0,\;l>-1/2,\;k<3l+7/2,
\]
as in~\cite{AR2u2}. This gives an equation for $\mu$, which can be rewritten
in terms of the function
\[
y(r)=\frac{e^{\mu(r)}}{E_0}.
\]
In this way the cut-off energy disappears as a free parameter of the problem and we
thus have the four free parameters $k,l,L_0$ and $y(0)$.
The structure of the static solutions obtained in~\cite{AR2u2} is as follows:

If $L_0=0$ the energy density can be strictly positive or vanish
at $r=0$ (depending on $l$) but it is always strictly positive sufficiently close
to $r=0$. Hence, the support of the matter is an interval $[0,R_1]$ with $R_1>0$,
and we call such states ball configurations. If $L_0>0$ the support is in an
interval $[R_0,R_1],\;R_0>0$, and we call such steady states for shells.

The value $y(0)$ determines how compact or relativistic the steady state is, and
the smaller values the more relativistic. For large values, recall $y(0)\leq 1/E_0$,
a pure shell or a pure ball configuration is obtained, cf.~Figure~\ref{fig1} for a pure shell.
Note that we depict the behavior of $\rho$ but we remark that the pressure 
terms behave similarly but the amplitudes of $p$ and $p_T$ can be very different, 
i.e., the steady states can be highly anisotropic.

\epubtkImage{fig20paper.png}{%
\begin{figure}[htbp]
\centerline{\includegraphics[scale=0.6]{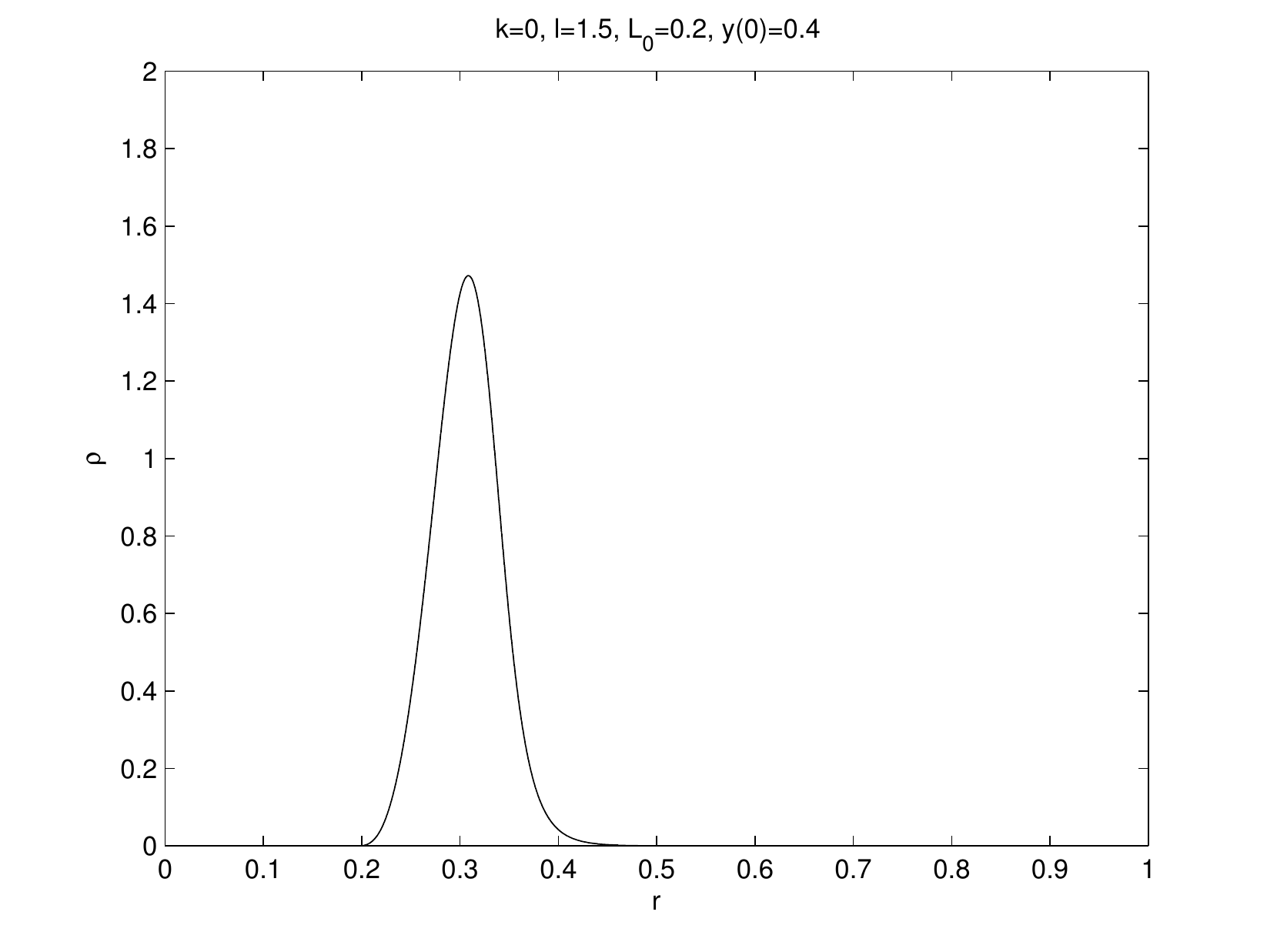}}
\caption{A pure shell}
\label{fig1}
\end{figure}
}

For moderate values of $y(0)$ the solutions have a distinct inner peak and a tail-like
outer peak, and by making $y(0)$ smaller more peaks appear, cf.~Figure~\ref{fig2} for 
the case of ball configurations.

\epubtkImage{fig2paper.png}{%
\begin{figure}[htbp]
\centerline{\includegraphics[scale=0.6]{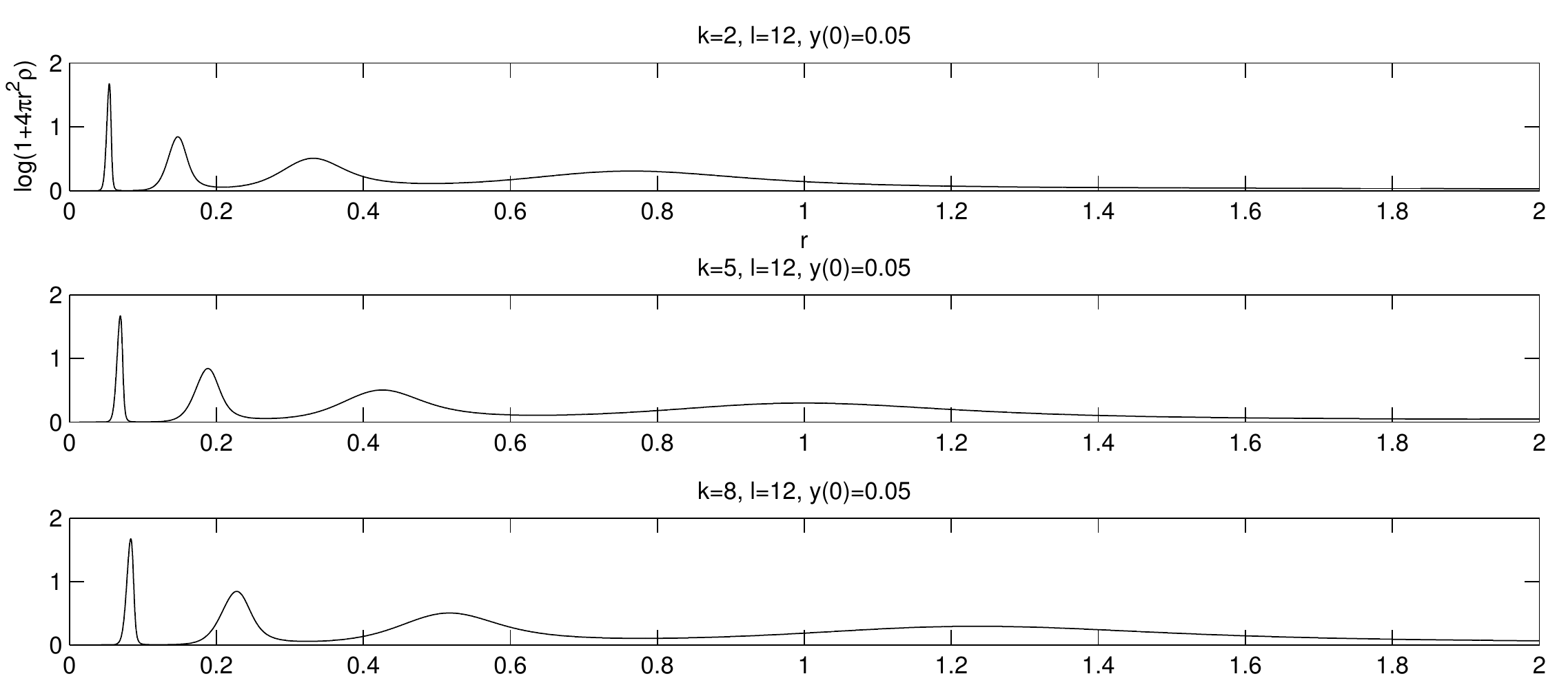}}
\caption{Multi-peaks of ball configurations, $L_0=0$}
\label{fig2}
\end{figure}
}

In the case of shells there is a similar structure but in this case the peaks
can either be separated by vacuum regions or by thin atmospheric regions as
in the case of ball configurations.
An example with multi-peaks, where some of the peaks are separated by
vacuum regions, is given in Figure~\ref{fig3}.

\epubtkImage{fig7paper.png}{%
\begin{figure}[htbp]
\centerline{\includegraphics[scale=0.6]{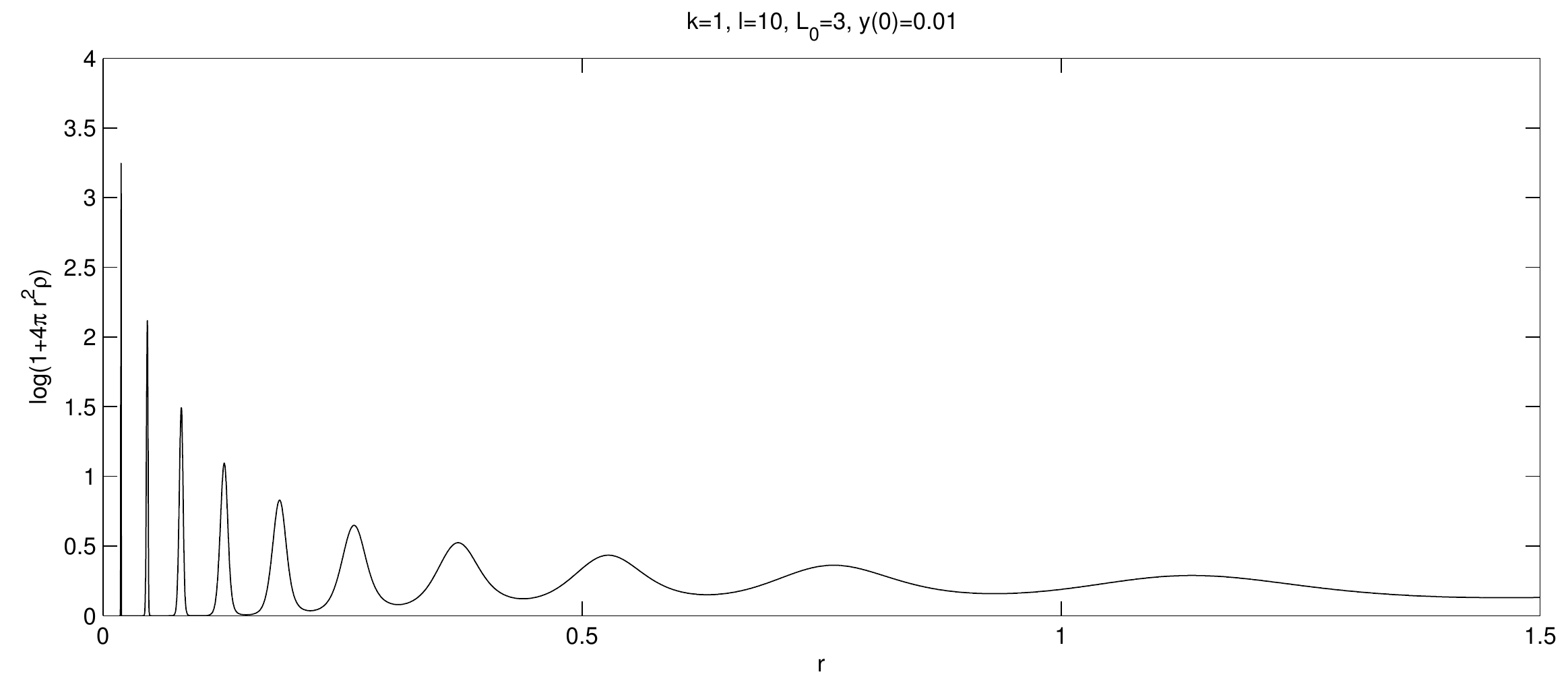}}
\caption{Multi-peaks of a shell}
\label{fig3}
\end{figure}
}

A different feature of the structure of static solutions is the issue of spirals.
For a fixed
ansatz of the density function $f$, there is a
one-parameter family of static
solutions, which are parameterized by $y(0)$. 
A natural question to ask is how the ADM mass $M$ and the radius
of the support $R$ change along such a family. By plotting
for each $y(0)$ the resulting values for 
$R$ and $M$ a curve is obtained, which reflects how radius
and mass are related along such a one-parameter family of steady states. This
curve has a spiral form, cf.~Figure~\ref{nonisospiral}. It is shown in~\cite{AR2u2}
that in the isotropic case, where $l=L_0=0$ the radius-mass curves always have
a spiral form.

\epubtkImage{spiraluniso.png}{%
\begin{figure}[htbp]
\centerline{\includegraphics[scale=0.6]{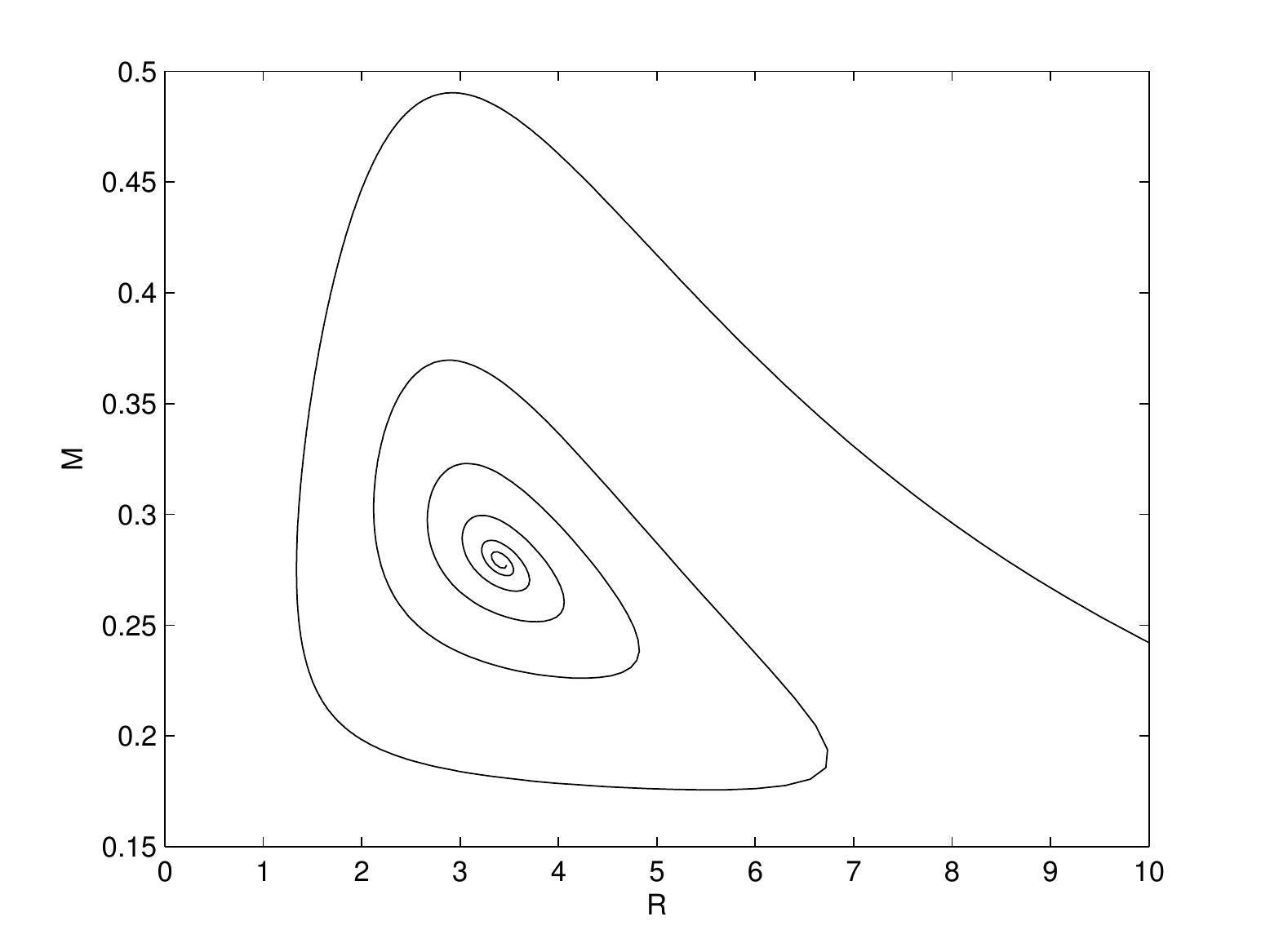}}
\caption{$(R,M)$ spiral for $k=0,\ l=10.5,\ L_0=0,\ 0.01 \leq y(0) \leq 0.99$}
\label{nonisospiral}
\end{figure}
}

\subsection{Buchdahl-type inequalities}

Another aspect of the structure of steady states
investigated numerically in~\cite{AR2u2} concerns
the {\it Buchdahl inequality}. If a steady state
has support in $[R_0,R_1]$, then the ADM mass $M$ of
the configuration is $M=m(R_1)$, where the quasi local
mass $m(r)$ is given by Equation (\ref{mass}) in Schwarzschild
coordinates.

In view of the Schwarzschild metric (\ref{Schwarz}), Schwarzschild 
asked already in 1916 the question: How large can
$2M/R$ possibly be? He gave the answer~\cite{Swu2} 
$2M/R\leq 8/9$ in the special case of the Schwarzschild interior solution, which has
constant energy density and isotropic pressure. In 1959
Buchdahl~\cite{Bu} extended his result to isotropic solutions for
which the energy density is non-increasing outwards and he showed that also
in this case
\begin{equation}
\label{Buchdahl}
\frac{2M}{R}\leq \frac89.
\end{equation}
This is sometimes called the Buchdahl inequality. 
Let us remark that the Buchdahl inequality can obviously be written as $M/R\leq 4/9$, 
but since it is the quantity $2M/R$, which appears in the Schwarzschild metric~(\ref{Schwarz}),
it is common to keep the form of Equation~(\ref{Buchdahl}).
A bound on $2M/R$ has an immediate observational
consequence since it limits the possible gravitational red shift
of a spherically-symmetric static object.

The assumptions made by Buchdahl are very restrictive. In particular,
the overwhelming number of the steady states
of the Einstein--Vlasov system have neither
an isotropic pressure nor a non-increasing energy density, but nevertheless $2M/R$
is always found to be less than $8/9$ in the numerical study~\cite{AR2u2}. 
Also for other matter models 
the assumptions are not satisfying. As 
pointed out by Guven and \'{O}~Murchadha~\cite{GMuu2}, neither of
the Buchdahl assumptions hold in a simple soap bubble and they do not
approximate any known topologically stable field configuration.
In addition, there are also several astrophysical models of stars, which are
anisotropic. Lem\^aitre~\cite{Leu2} proposed a model of an anisotropic star
already in 1933, and Binney and Tremaine~\cite{BTu2} explicitly allow
for an anisotropy coefficient. Hence, it is an important question to
investigate bounds on $2M/R$ under less restrictive assumptions. 

In~\cite{A4u2} it is shown that for any static solution of the
spherically-symmetric Einstein equation, not necessarily of the
Einstein--Vlasov system, for which $p\geq 0$, and
\begin{equation}
\label{Omega}
p+2p_T\leq\Omega\rho,
\end{equation}
the following inequality holds
\begin{equation}
\label{sbound}
\frac{2m(r)}{r}\leq\frac{(1+2\Omega)^2-1}{(1+2\Omega)^2}.
\end{equation}
Moreover, the inequality is sharp and sharpness is obtained uniquely by an infinitely
thin shell solution. Note in particular that for Vlasov matter $\Omega=1$ and
that the right-hand side then equals $8/9$ as in the Buchdahl inequality.
An alternative proof was given in~\cite{KSu2} and their method applies to a larger
class of conditions on $\rho,p$ and $p_T$ than the one given in Equation~(\ref{Omega}). On
the other hand, the result in~\cite{KSu2} is weaker than
the result in~\cite{A4u2} in the sense that the latter method
implies that the steady state that saturates the inequality is
unique; it is an infinitely thin shell.
The studies~\cite{A4u2,KSu2} are of general character and in particular
it is not shown that solutions exist to the \textit{coupled} Einstein-matter system, which can saturate the inequality. For instance, it is natural
to ask if there are solutions of the Einstein--Vlasov system, which have $2m/r$
arbitrarily close to $8/9$. This
question is given an affirmative answer in~\cite{A3u2}, where in
particular it is shown that arbitrarily thin shells exist, which are
regular solutions of the spherically-symmetric Einstein--Vlasov system. Using 
the strategy in~\cite{A2u2} it follows that 
\[
\sup_r\frac{2m(r)}{r}\to\frac89,
\]
in the limit when the shells become infinitely thin. 

The question of finding an upper bound on $2M/R$ can be extended to charged
objects and to the case with a positive cosmological constant.
The spacetime outside a spherically-symmetric charged object is given by
the Reissner--Nordstr\"{o}m metric
\[
ds^2=-\big(1-\frac{2M}{r}+\frac{Q}{r^2}\big)\,dt^2
+\big(1-\frac{2M}{r}+\frac{Q}{r^2}\big)^{-1}\,dr^2
+r^2(d\theta^2+\sin^2{\theta}\,d\phi^2),
\]
where $Q$ is the total charge of the object. The quantity $1-\frac{2M}{r}+\frac{Q}{r^2}$
is zero when $r_\pm=M\pm\sqrt{M^2-Q^2}$, and $r_\pm$ is called the inner and outer horizon
respectively of a
Reissner--Nordstr\"{o}m black hole. A Buchdahl type inequality gives a lower
bound of the area radius of a static object and this radius is thus often called
the critical stability radius. It is shown in~\cite{A5u2} that
a spherically-symmetric static solution of the Einstein--Maxwell system for
which $p\geq 0,\; p+2p_T\leq \rho$, and $Q<M$ satisfy
\begin{equation}
\label{DAineq}
\sqrt{M}\leq\frac{\sqrt{R}}{3}+\sqrt{\frac{R}{9}+\frac{Q^2}{3R}}.
\end{equation}
Note, in particular,
that the inequality holds for solutions of the Einstein--Vlasov--Maxwell system,
since the conditions above are always satisfied in this case.
This inequality~(\ref{DAineq}) implies that the stability radius is outside
the outer horizon of a Reissner--Nordstr\"{o}m black hole. In~\cite{GRu2} the relevance 
of an inequality of this kind on aspects in black-hole physics is discussed. In contrast
to the case without charge, the saturating solution is not unique. An infinitely thin
shell solution does saturate the inequality~(\ref{DAineq}), but numerical
evidence is given in~\cite{AERu2} that there is also another type of solution, which
saturates the inequality for which the inner and outer horizon coincide. 

The study in~\cite{ABu2} is concerned with the non-charged situation when a positive
cosmological constant $\Lambda$ is included. The following inequality is derived
\begin{equation*}
\label{ineqlambda}
  \frac{M}{R}\leq\frac29-\frac{\Lambda R^2}{3}+\frac29 \sqrt{1+3\Lambda R^2},
\end{equation*}
for solutions for which $p\geq 0,\; p+2p_T\leq \rho$, and $0\leq\Lambda R^2\leq 1$.
In this situation, the question of sharpness is essentially open. An infinitely
thin shell solution does not generally saturate the inequality but does so in the two
degenerate situations $\Lambda R^2=0$ and $\Lambda R^2=1$. In the latter case
there is a constant density solution, and the exterior spacetime is the Nariai solution,
which saturates the inequality and the saturating solution is thus non-unique.
In this case, the cosmological horizon and the black hole horizon coincide, which is
in analogy with the charged situation described above where the inner and outer
horizons coincide when uniqueness is likely lost.

\subsection{Stability}

An important problem is the question of the stability of spherically-symmetric steady states. 
At present, there are almost no theoretical results on the stability of
the steady states of the Einstein--Vlasov system.
Wolansky~\cite{Wo} has applied the energy-Casimir method and obtained
some insights, but the theory is much less developed than 
in the Vlasov--Poisson case and the stability problem is essentially open.
The situation is very different for the Vlasov--Poisson system,
and we refer to~\cite{Rn1u2} for a review on the results in this case. 

However, there are numerical studies~\cite{AR1u2,Iu2,ST1u2} on the stability of 
spherically-symmetric steady
states for the Einstein--Vlasov system. The latter two studies concern isotropic steady
states, whereas the first, in addition, treats anisotropic steady states. Here we 
present the conclusions of~\cite{AR1u2}, emphasizing that these agree
with the conclusions in~\cite{ST1u2,Iu2} for isotropic states. 

To allow for trapped surfaces, maximal-areal coordinates are used, i.e.,
the metric is written in the following form in~\cite{AR1u2}
\[
ds^{2}=-(\alpha^2-a^2\beta^2)dt^2+2a^2\beta dtdr+a^2 dr^2
+ r^2\left(d\theta^2 + \sin^2\theta\, d\phi^2\right).
\]
Here the metric coefficients $\alpha, \beta$, and $a$ depend on
$t\in \mathbb{R}$ and $r\geq 0$,
$\alpha$ and $a$ are positive, and the polar angles $\theta\in[0,\pi]$
and $\phi\in[0,2\pi]$ parameterize the unit sphere.
Thus, the radial coordinate $r$ is the area radius.
A maximal gauge condition is then imposed, which means that
each hypersurface of constant $t$ has vanishing mean curvature.
The boundary conditions, which guarantee asymptotic flatness and a regular center, 
are given by
\begin{equation}
a(t,0)=a(t,\infty)=\alpha(t,\infty)=1.\label{bdryc}
\end{equation}
Steady states are numerically constructed, and these are then perturbed 
in order to investigate the stability. More precisely, to construct the steady states
the polytropic ansatz is used, cf. Section~\ref{existssss},
\begin{equation}
f(r,w,L)=\Phi(E,L)=(E_0-E)^k_{+}(L-L_0)_{+}^l.
\end{equation}
%
By specifying values on $E_{0},\, L_{0}$ and $\alpha(0)$ steady states
are numerically constructed.
The distribution function $f_s$ of the steady state
is then multiplied by an amplitude $A$, so that a new, perturbed
distribution function is obtained. This is then used as initial
datum in the evolution code. We remark that also 
other types of perturbations are analyzed in~\cite{AR1u2}.

For $k$ and $l$ fixed each steady state is characterized
by its
central red shift $Z_{c}$ and its fractional binding energy
$E_{b}$, which are defined by
\[
Z_{c}=\frac{1}{\alpha(0)}-1,\;\; E_{b}=\frac{e_b}{M_{0}},
\text{ where } e_{b}=M_{0}-M.
\]
%
Here
\[
M_0=4\pi^2\int_{0}^{\infty}\int_{-\infty}^{\infty}\int_{0}^{\infty}
a(t,r)f(t,r,w,L)\,dL\,dw\,dr
\]
is the total number
of particles, which, since all particles
have rest mass one, equals the rest mass of the system. $M$ is the ADM mass
given by
\[
M=\int_{0}^{\infty}\left(4\pi \rho(t,r)+\frac{3}{2}
\kappa^2(t,r)\right) r^2 dr,
\]
where $\kappa=\beta/r\alpha$. Both $M_0$ and $M$ are conserved quantities.
The central redshift is the redshift of a photon emitted from the
center and received at infinity, and the binding energy $e_b$ is
the difference of the rest mass and the ADM mass.
In Figure~\ref{fig5} and Figure~\ref{fig6} the relation between
the fractional binding energy and the central redshift is given
for two different cases.

\epubtkImage{case1ebzc3f001.png}{%
\begin{figure}[htbp]
\centerline{\includegraphics[scale=0.6]{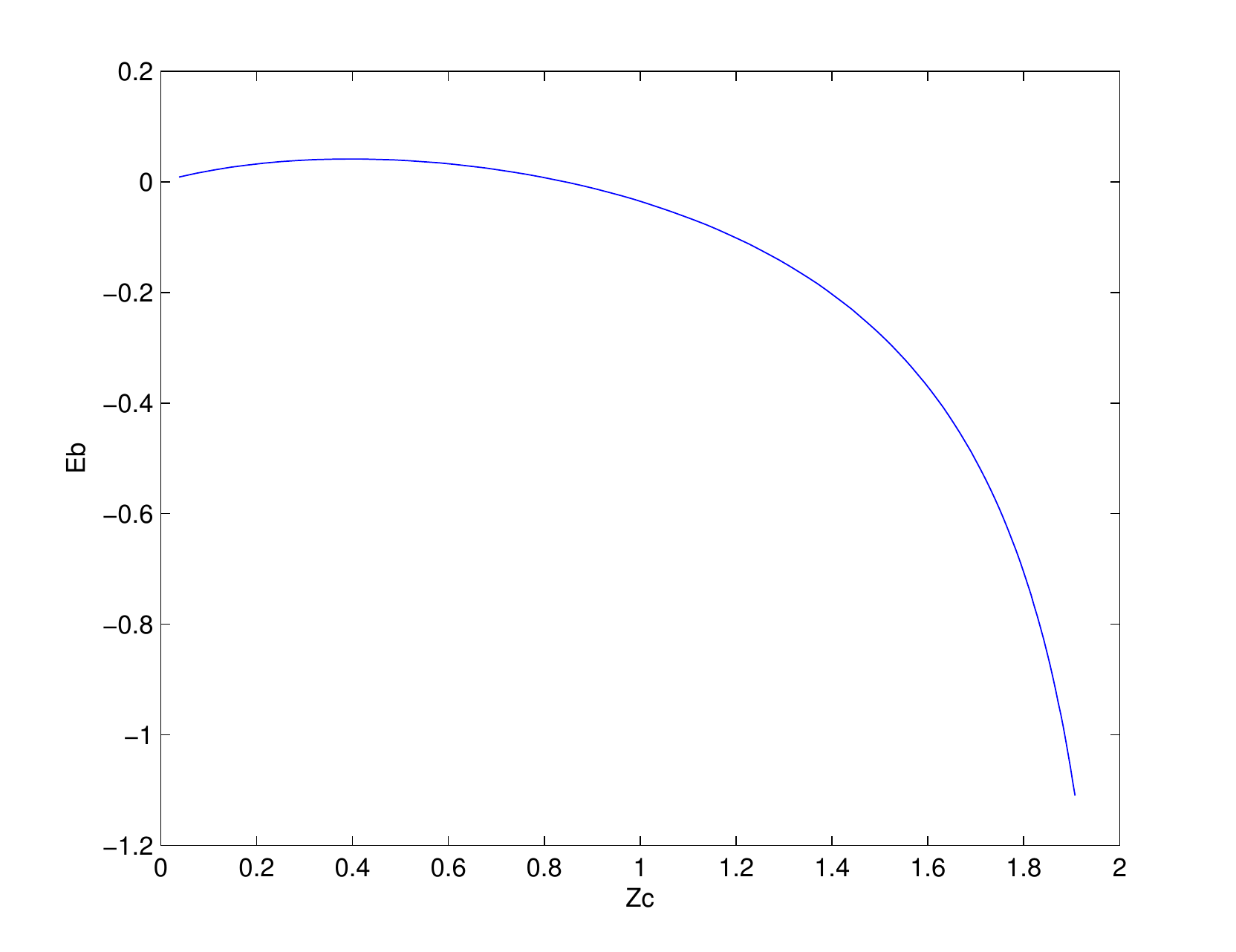}}
\caption{$k=0,\, l=0,\, L_0=0.1$}
\label{fig5}
\end{figure}
}

\epubtkImage{case4ebzc3f001.png}{%
\begin{figure}[htbp]
\centerline{\includegraphics[scale=0.6]{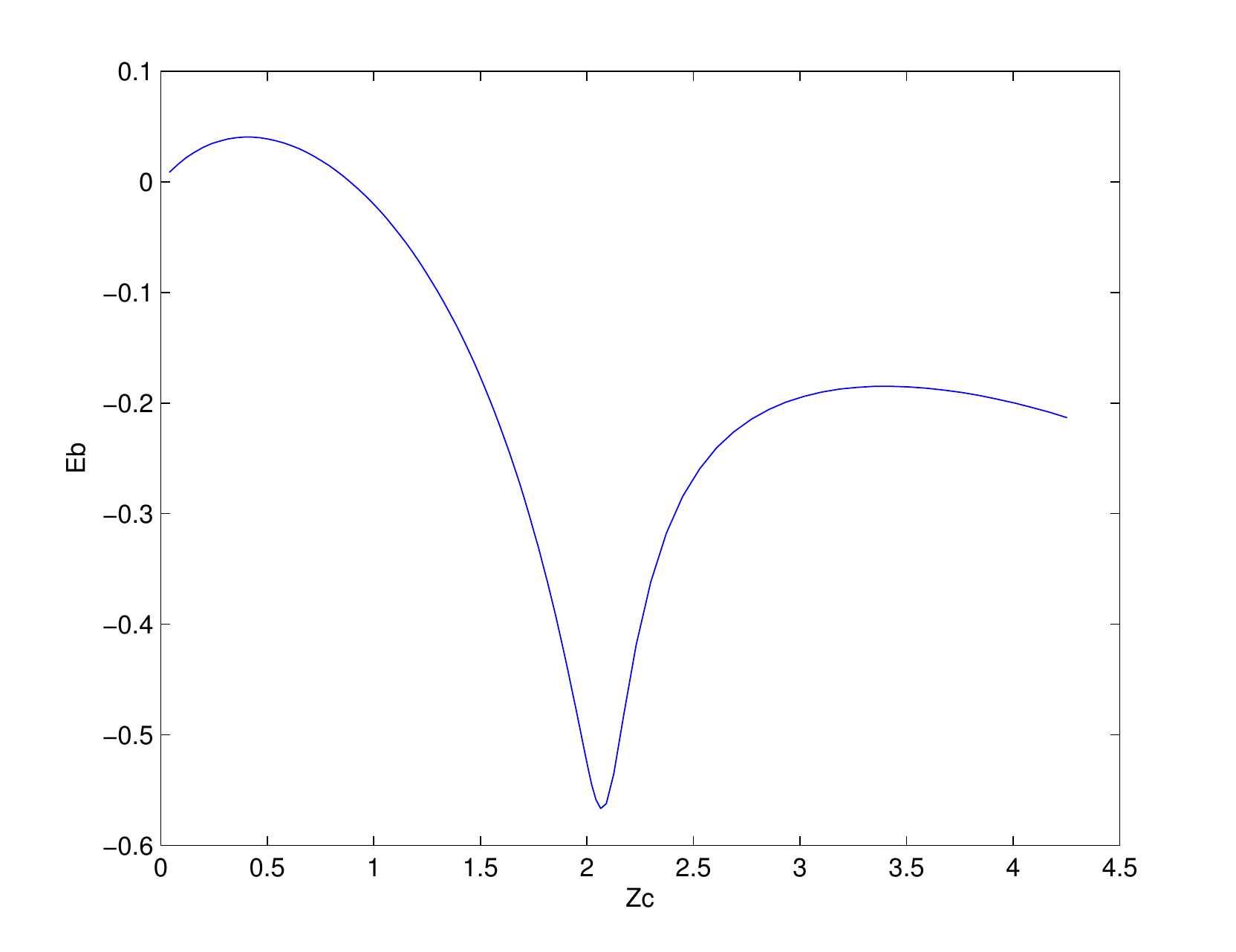}}
\caption{$k=0,\, l=3/2,\, L_0=0.1$}
\label{fig6}
\end{figure}
}

The relevance of these concepts for the stability properties of
steady states was first discussed by Zel'dovich and
Podurets~\cite{ZPu2}, who argued that it should be possible to diagnose
the stability from binding energy considerations. Zel'dovich and
Novikov~\cite{ZNu2} then conjectured that the binding energy maximum
along a steady state sequence signals the onset of instability.

The picture that arises from the simulations in~\cite{AR1u2} is summarized
in Table~\ref{tab:1}. Varying the parameters $k,l$ and $L_0$ give rise to
essentially the same tables, cf.~\cite{AR1u2}.

\begin{table}[h]
\caption{$k=0$ and $l=1/2$.}
\label{tab:1}
\centering
\begin{tabular}{rrcc}
\toprule
$Z_c$ & $E_b$ & $A<1$ & $A>1$ \\ 
\midrule
0.21  & 0.032 & stable & stable \\
0.34  & 0.040 & stable & stable \\
0.39  & 0.040 & stable & stable \\
0.42  & 0.041 & stable & unstable \\
0.46  & 0.040 & stable & unstable \\
0.56  & 0.036 & stable & unstable \\
0.65  & 0.029 & stable & unstable \\
0.82  & 0.008 & stable & unstable \\
0.95  & --0.015 & unstable & unstable \\
1.20  & --0.078 & unstable & unstable \\ 
\bottomrule
\end{tabular}
\end{table}

If we first consider perturbations with $A>1$, it is found
that steady states with small values on $Z_c$ (less than
approximately 0.40 in this case) are
stable, i.e., the perturbed solutions
stay in a neighbourhood of the static solution.
A careful investigation of the perturbed solutions 
indicates that they oscillate in a periodic way.
For larger values of $Z_c$ the evolution leads to the formation
of trapped surfaces and collapse to
black holes.
Hence, for perturbations with $A>1$ the value of $Z_c$ alone seems
to determine the stability features of the steady states.
Plotting $E_b$ versus $Z_c$ with higher resolution, cf.~\cite{AR1u2},
gives support to the conjecture by Novikov and Zel'dovich mentioned
above that the maximum of $E_b$ along a sequence of steady states
signals the onset of instability.

The situation is quite different for perturbations with $A<1$.
The crucial quantity in this case is the fractional binding energy
$E_b$. Consider a steady state with $E_b >0$
and a perturbation with $A<1$ but close to 1
so that the fractional binding energy remains positive. The perturbed solution 
then drifts outwards, turns back and reimplodes, and comes close to its 
initial state, and then continues to expand and reimplode and thus
oscillates, cf.~Figure~\ref{fig7}.

\epubtkImage{a099T90box.png}{%
\begin{figure}[htbp]
\centerline{\includegraphics[scale=0.9]{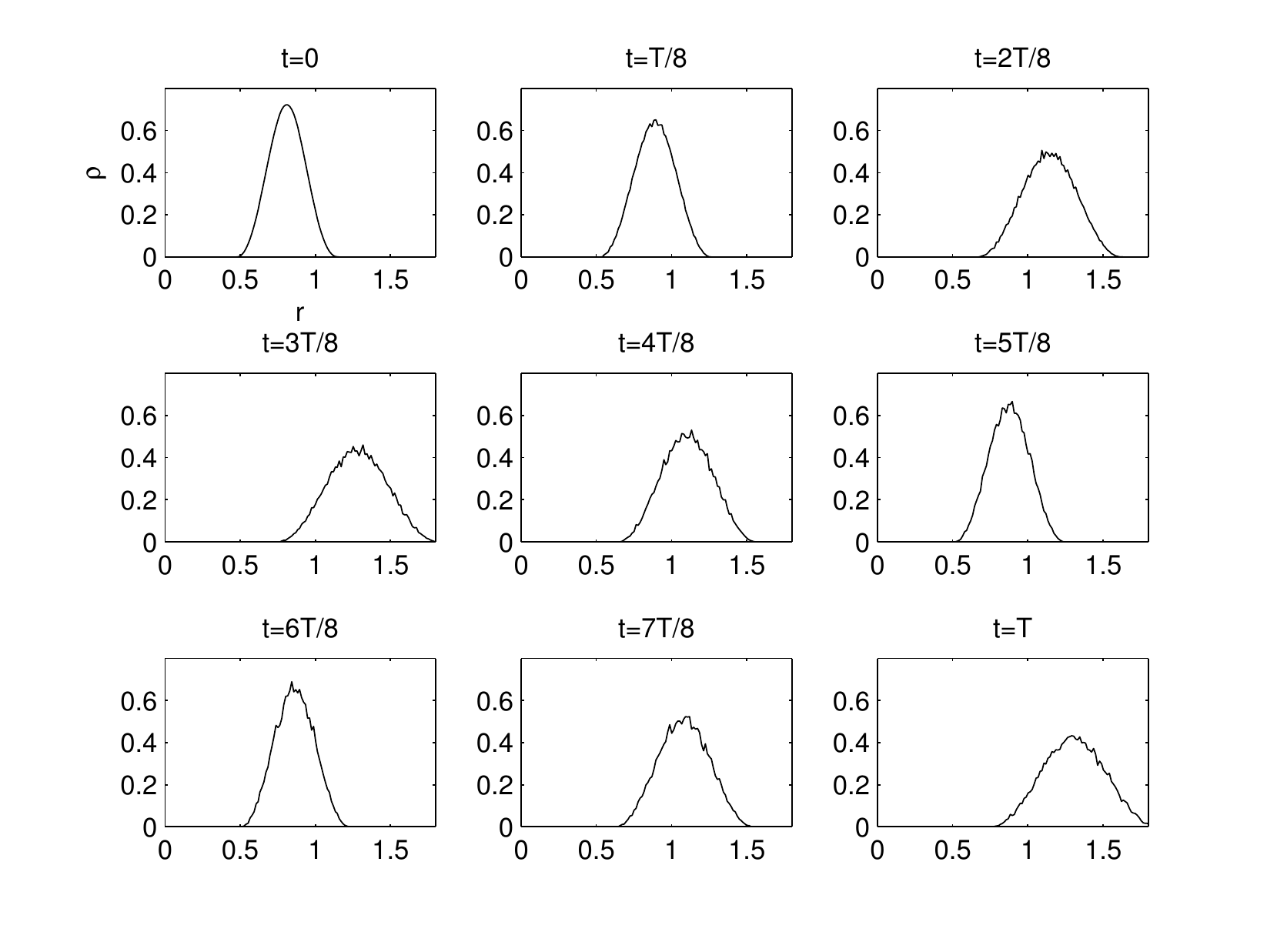}}
\caption{$Z_c=0.47,\ E_b=0.04,\ A=0.99,\ T=90.0$}
\label{fig7}
\end{figure}
}

In~\cite{ST1u2} it is stated (without proof) that if $E_b>0$
the solution must ultimately reimplode and the simulations
in~\cite{AR1u2} support that it is true. For negative values of $E_b$, the
solutions with $A<1$ disperse to infinity. 

A simple analytic argument is given in~\cite{AR1u2}, which relates
the question, whether a solution disperses or not. It is shown
that if a shell solution has an expanding vacuum region
of radius $R(t)$ at the center with $R(t) \to \infty$
for $t \to \infty$, i.e., the solution disperses in a strong sense,
then necessarily $M_0 \leq M$, i.e., $E_b \leq 0$.


\subsection{Existence of axisymmetric static solutions}

As we have seen above, a broad variety
of static solutions of the Einstein--Vlasov system
has been established,
all of which share the restriction that they are spherically
symmetric. The recent investigation~\cite{AKR4u2}
removes this restriction and proves the existence of
static solutions of the Einstein--Vlasov system,
which are axially symmetric but not spherically symmetric.
From the applications
point of view this symmetry assumption is more ``realistic''
than spherical symmetry, and from the mathematics point of view
the complexity of the Einstein field equations increases drastically
if one gives up spherical symmetry. Before discussing this result, 
let us mention that similar 
results have been obtained for two other matter models. In the case of a
perfect fluid, Heilig showed the existence of axisymmetric stationary
solutions in~\cite{Hu2}. These solutions have non-zero angular momentum since 
static solutions are necessarily spherically symmetric. In this respect
the situation for elastic matter is more similar to Vlasov matter.
The existence of static axisymmetric solutions of elastic matter,
which are not spherically symmetric, was proven in~\cite{ABS1u2}.
Stationary solutions with rotation were then established in~\cite{ABS2u2}.

Let us now briefly discuss the method of proof in~\cite{AKR4u2}, which relies on
an application of the implicit function theorem. Also, 
the proofs in~\cite{Hu2, ABS1u2, ABS2u2} make use of the implicit function theorem, but 
apart from this fact the methods are quite different.

The set-up of the problem in~\cite{AKR4u2} follows the work of Bardeen~\cite{Bu2},
where the metric is written
in the form
\begin{equation}
\label{metric_ax}
ds^2=-c^2 e^{2\nu/c^2} dt^2 + e^{2\mu} d\rho^2 + e^{2\mu} dz^2+
\rho^2 B^2 e^{-2\nu/c^2} d\varphi^2
\end{equation}
for functions $\nu, B, \mu$ depending on $\rho$ and $z$, where
$t\in\R,\,\rho \in [0,\infty[,\,z\in \R$ and $\varphi\in [0,2 \pi]$.
The Killing vector fields $\partial_t$ and $\partial_\phi$ correspond to the stationarity 
and axial symmetry of the spacetime.
Solutions are obtained by perturbing off spherically symmetric steady states
of the Vlasov--Poisson system via the implicit function theorem and the reason
for writing $\nu/c^2$ in the metric, instead of $\nu$,
is that $\nu$ converges to the Newtonian potential
$U_N$ of the steady state in the limit $c\to \infty$.
Asymptotic flatness is expressed by
the boundary conditions
\begin{equation}
\label{bc_infinity}
\lim_{|(\rho,z)| \to \infty} \nu(\rho,z) =
\lim_{|(\rho,z)| \to \infty} \mu(\rho,z) = 0,\
\lim_{|(\rho,z)| \to \infty} B(\rho,z) = 1.
\end{equation}
In addition the solutions are required to be locally flat
at the axis of symmetry, which implies the condition
\begin{equation}
\label{bc_axis}
\nu(0,z)/c^2 + \mu(0,z) = \ln B(0,z),\ z\in \R .
\end{equation}
Let us now recall from Section~\ref{existssss} the strategy to 
construct static solutions by using an ansatz of the form
\begin{equation}
  f(x,v)=\Phi(E,L)\nonumber,
\end{equation}
where $E$ and $L$ are conserved quantities along characteristics.
Due to the symmetries of the metric~(\ref{metric_ax}) the following quantities
are constant along geodesics:
\begin{eqnarray}
E
&:=& - g(\partial/\partial t, p^a) = c^2 e^{2\nu/c^2} p^0 \nonumber\\
&=&
c^2 e^{\nu/c^2} \sqrt{1+c^{-2}\left(e^{2\mu} (p^1)^2 + e^{2\mu} (p^2)^2 +
\rho^2 B^2 e^{-2\nu/c^2} (p^3)^2 \right)},\quad \label{Edef}\\
L
&:=&
g(\partial/\partial \varphi, p^a) =
\rho^2\,B^2 e^{-2\nu/c^2} p^3. 
\label{Ldef}
\end{eqnarray}
Here $p^a$ are the canonical momenta. 
$E$ can be thought of as a local or particle energy and $L$
is the angular momentum of a particle with respect to the axis
of symmetry.
For a sufficiently regular $\Phi$ the ansatz function $f$
satisfies the Vlasov equation and upon insertion of this ansatz into the definition
of the energy momentum tensor~(\ref{enermomts}) the latter
becomes a functional $T_{\alpha \beta} =  T_{\alpha \beta} (\nu,B,\mu)$
of the unknown metric functions
$\nu,B,\mu$. It then remains to solve the Einstein equations
with this energy momentum tensor as right-hand side.
The Newtonian limit of the Einstein--Vlasov system is the Vlasov--Poisson system
and the strategy in~\cite{AKR4u2} is to perturb off spherically symmetric steady
states of the Vlasov--Poisson system via the implicit function theorem
to obtain axisymmetric solutions. Indeed, the main result of~\cite{AKR4u2}
specifies conditions on the ansatz function
$\Phi$ such that a two parameter ($\gamma$ and $\lambda$) family of axially-symmetric
solutions of the Einstein--Vlasov system passes through the corresponding
spherically symmetric, Newtonian steady state, whose ansatz function we denote by $\phi$. 
The parameter
$\gamma = 1/c^2$ turns on general relativity and the parameter $\lambda$
turns on the dependence on $L$.
Since $L$ is not invariant under arbitrary rotations about
the origin the solution is not spherically symmetric if
$f$ depends on $L$. It should also be mentioned that although $\gamma$ is a priori
small, which means that $c$ is large,
the scaling symmetry of the Einstein--Vlasov system can be used to obtain solutions
corresponding to the physically correct value of $c$.
The most striking condition on the ansatz function $\Phi$, or rather on the ansatz 
function $\phi$ of the corresponding Vlasov--Poisson system, needed to carry out the proof
is that it must satisfy 
\[
6+4\pi r^2 a_N(r)>0,\quad r\in [0, \infty[,
\]
where
\[
a_N(r):=\int_{\R^3}\phi'\Big(\frac{1}{2}\,|v|^2+U_N(r)\Big)\,dv.
\]
An important argument in the proof is indeed to justify that there are steady states
of the Vlasov--Poisson system satisfying this condition.

It is of course desirable to extend the result in~\cite{AKR4u2} to
stationary solutions with rotation. Moreover, the deviation from
spherically symmetry of the solutions in~\cite{AKR4u2} is small and an
interesting open question is the existence of disk-like models of
galaxies. In the Vlasov--Poisson case this has been shown in~\cite{FRu2}.


\section{Acknowledgements}

I would like to thank Alan Rendall for helpful suggestions.

\newpage


\bibliography{article}

\end{document}